\documentclass[apj]{emulateapj}

\usepackage{times}
\usepackage{graphicx}

\shortauthors{Brooks et al.}
\shorttitle{Active region outflows}

\begin{document}


\title{The drivers of active region outflows into the slow solar wind}

\author{David H. Brooks\altaffilmark{1,*}, Amy R. Winebarger\altaffilmark{2}
Sabrina Savage\altaffilmark{2}, Harry P. Warren\altaffilmark{3}, Bart De Pontieu\altaffilmark{4,5}, Hardi Peter\altaffilmark{6}, 
Jonathan W. Cirtain\altaffilmark{7}, Leon Golub\altaffilmark{8}, Ken Kobayashi\altaffilmark{2}, Scott W. McIntosh\altaffilmark{9}, 
David McKenzie\altaffilmark{2}, Richard Morton\altaffilmark{10}, Laurel Rachmeler\altaffilmark{2}, Paola Testa\altaffilmark{8}, 
Sanjiv Tiwari\altaffilmark{4, 11}, Robert Walsh\altaffilmark{12} }

\affiliation{\altaffilmark{1}College of Science, George Mason University, 4400 University Drive,
  Fairfax, VA 22030}

\altaffiltext{*}{Current address: Hinode Team, ISAS/JAXA, 3-1-1 Yoshinodai, Chuo-ku, Sagamihara,
  Kanagawa 252-5210, Japan}

\affiliation{\altaffilmark{2}NASA Marshall Space Flight Center, 320 Sparkman Dr. NW, Huntsville, AL 35805, USA}

\affiliation{\altaffilmark{3}Space Science Division, Naval Research Laboratory, Washington DC 20375, USA}

\affiliation{\altaffilmark{4}Lockheed Martin Solar and Astrophysics Laboratory, 3251 Hanover St., Palo Alto, CA 94304, USA}

\affiliation{\altaffilmark{5}Institute of Theoretical Astrophysics, University of Oslo, P.O. Box 1029 Blindern, NO-0315 Oslo, Norway}

\affiliation{\altaffilmark{6}Max Planck Institute for Solar System Research, 37077 G\"{o}ttingen, Germany}

\affiliation{\altaffilmark{7}BWX Technologies, Inc., 800 Main St \#400, Lynchburg, VA 24504, USA}

\affiliation{\altaffilmark{8}Harvard-Smithsonian Center for Astrophysics, 60 Garden Street, Cambridge, MA 02138, USA}

\affiliation{\altaffilmark{9}High Altitude Observatory, National Center for Atmospheric Research, P.O. Box 3000, Boulder, CO 80307, USA}

\affiliation{\altaffilmark{10}Mathematics, Physics and Electrical Engineering, Northumbria University, Newcastle Upon Tyne NE1 8ST, UK}

\affiliation{\altaffilmark{11}Bay Area Environmental Research Institute, NASA Research Park, Moffett Field, CA 94035, USA,}

\affiliation{\altaffilmark{12}Jeremiah Horrocks Institute, University of Central Lancashire, Preston, PR1 2HE, UK}


\begin{abstract}
Plasma outflows from the edges of active regions have been suggested as a possible source of the slow solar wind. Spectroscopic measurements show that these outflows have an enhanced elemental composition, which is a distinct signature of the slow wind. Current spectroscopic observations, however, do not have sufficient spatial resolution to distinguish what structures are being measured or to determine the driver of the outflows. The High-resolution Coronal Imager (Hi-C) flew on a sounding rocket in May, 2018, and observed areas of active region outflow at the highest spatial resolution ever achieved (250\,km). Here we use the Hi-C data to disentangle the outflow composition signatures observed with the Hinode satellite during the flight. We show that there are two components to the outflow emission: a substantial contribution from expanded plasma that appears to have been expelled from closed loops in the active region core, and a second contribution from dynamic activity in active region plage, with a composition signature that reflects solar photospheric abundances. The two competing drivers of the outflows may explain the variable composition of the slow solar wind.
\end{abstract}



\section{introduction}
\label{introduction}
The source of the slow ($\sim$400\,km s$^{-1}$) solar wind that fills the heliosphere remains elusive, and several possibilities have been suggested and debated \citep{abbo_etal2016}. One promising candidate during periods of high solar activity is outflows from the edges of active regions \citep{sakao_etal2007, harra_etal2008, doschek_etal2008}. At solar maximum, at least some fraction of the mass supply to the slow wind appears to originate low down in the solar atmosphere in these outflows \citep{sakao_etal2007, brooks_etal2015}, and is often able to escape into interplanetary space on open magnetic field lines \citep{sakao_etal2007, harra_etal2008, doschek_etal2008}. While the composition of the fast ($>$700\,km s$^{-1}$) solar wind largely reflects solar photospheric abundances, the slow wind shows much more variability. The composition can also be close to photospheric, but is often enhanced above those levels by factors of 2-4 \citep{meyer_1985, vonsteiger_etal2000, stakhiv_etal2016} due to the first ionization potential (FIP) effect; where the plasma is enriched with elements of low ($<$10\,eV) FIP \citep{pottasch_1963, feldman_1992}.  An enhanced composition, similar to the closed-field solar corona, is a distinct signature of the slow solar wind, and has been detected in active region outflows using observations from the EUV Imaging Spectrometer \citep[EIS,][]{culhane_etal2007a} on Hinode \citep{brooks&warren_2011}.

\begin{figure*}[t!]
  \centerline{%
    \includegraphics[width=1.0\textwidth]{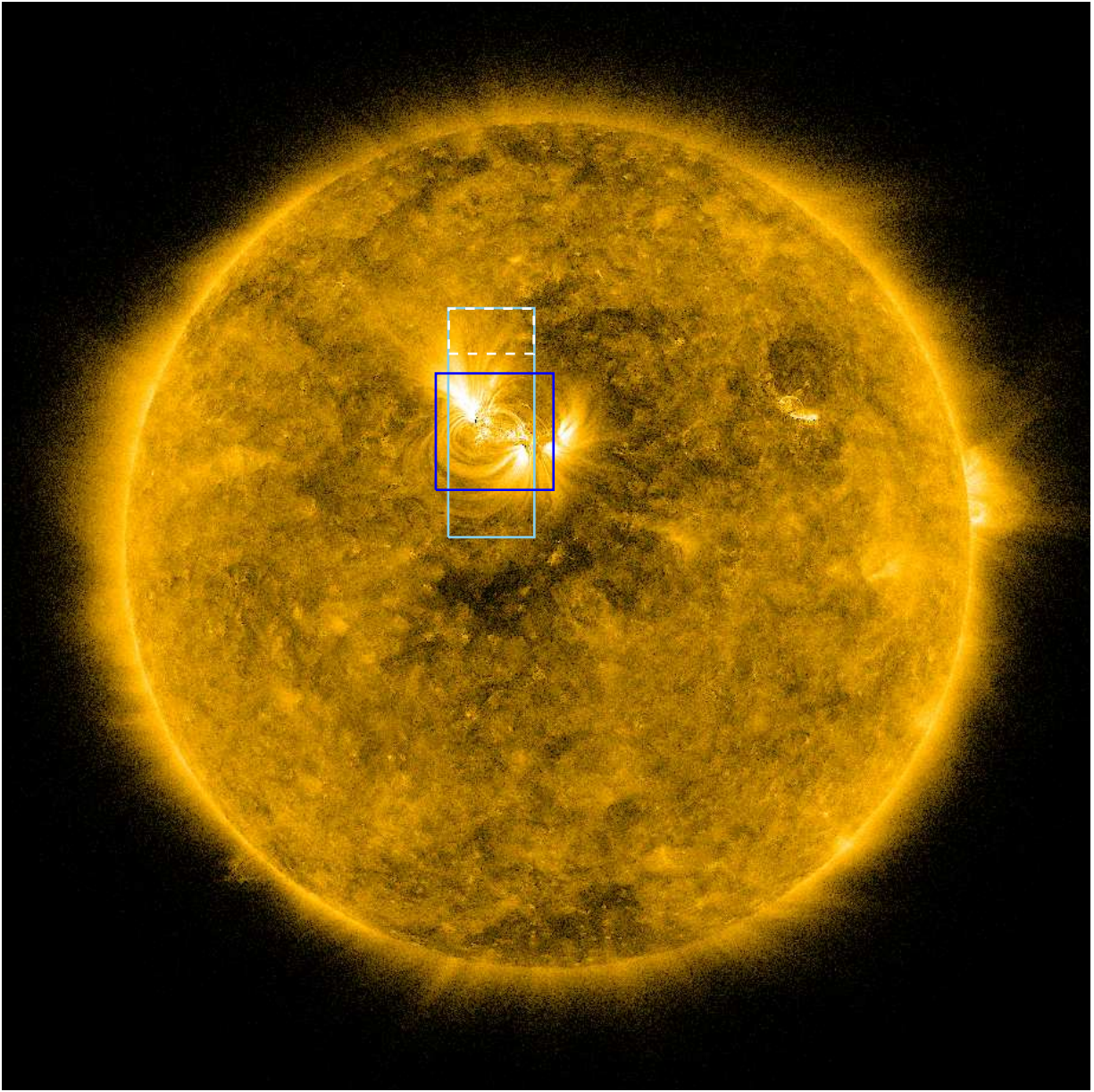}} %
  \caption{SDO/AIA 171\,\AA\, filter image of the solar disk on 2018, May 29, during the Hi-C flight.
The blue box shows the Hi-C field-of-view (FOV) from Fig. \ref{fig:fig3}, and the sky blue 
colored box shows the EIS FOV for the observations taken during the rocket flight.
The white dashed box shows the area used to determine
the rest wavelength of the Fe XII 202.044\,\AA\, line. This is then used to calculate relative Doppler velocities (see discussion in section \ref{velocity}). }
  \label{fig:fig1}
\end{figure*}

\begin{figure*}[t!]
  \centerline{%
    \includegraphics[width=1.0\textwidth]{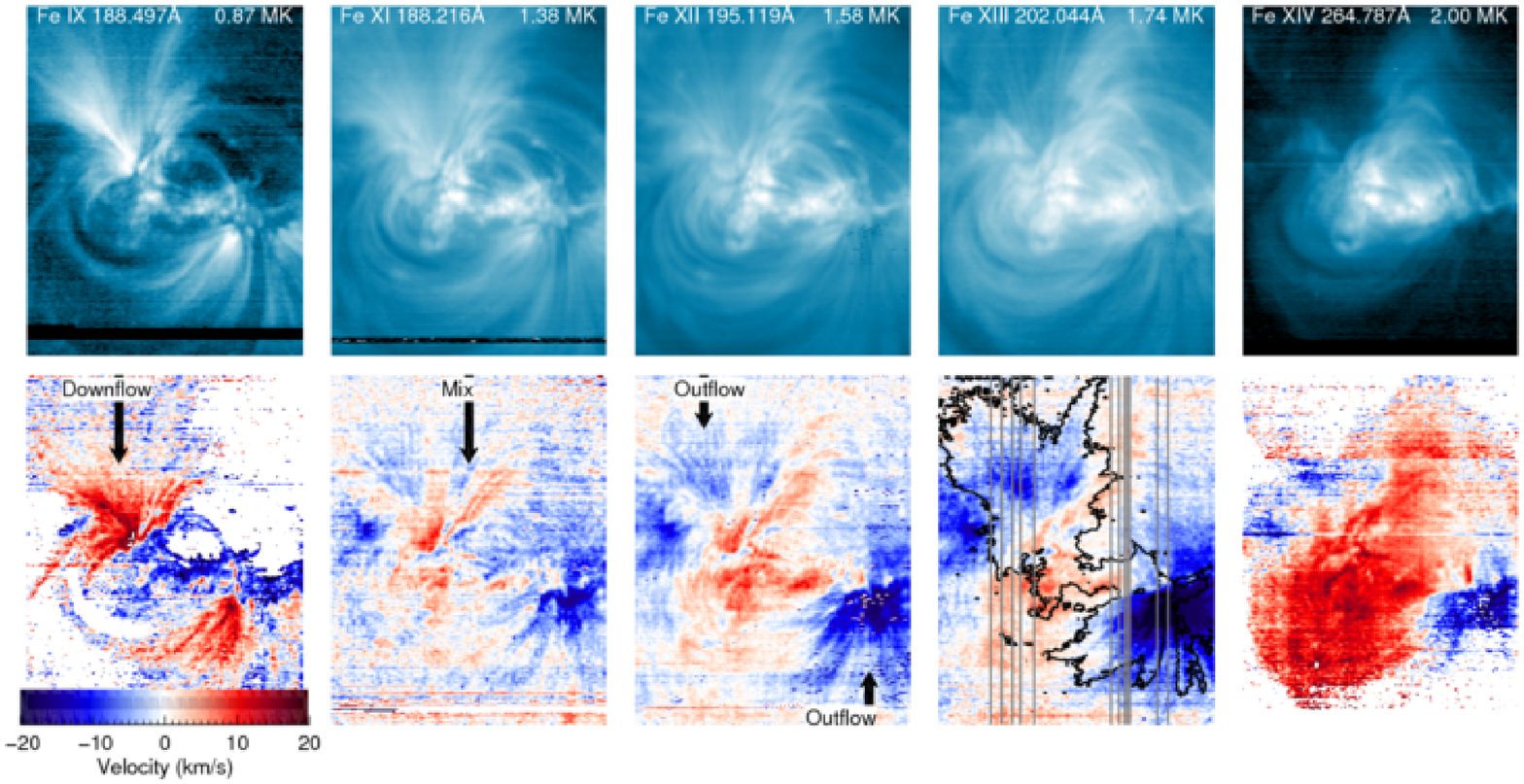}} %
  \caption{Hinode/EIS observations of AR 12712 on May 29, 2018.
Top row: images formed from Fe line emission at several wavelengths (given in the legends) obtained by fitting Gaussian functions to the spectral line profiles. The formation temperatures of the images are shown and increase left to right from 0.87\,MK to 2\,MK. Bottom row: Doppler velocity maps derived from the same spectral fits (as described in Section \ref{velocity}). Blue/red indicates areas of upflow/downflow (towards/away from the observer). We have overlaid contours of the Fe IX 188.497\,\AA\, intensity (top left image) on top of the Fe XIII 202.044\,\AA\, velocity map to show the relationship between the upflows at that temperature and the features seen at approximately the same temperature as Hi-C. We also show the approximate locations of the EIS slit positions in Fig. \ref{fig:fig3} for cross-reference (grey). These have been rotated back from the time of the observations to the time when the EIS slit reached the center of the images in this Figure. }
  \label{fig:fig2}
\end{figure*}

While the exact contribution of active region outflows to the slow wind is still under debate, they are scientifically interesting in themselves as they form part of the basic structure of active regions \citep{delzanna_2008}. Active regions are typically composed of a hot core emitting at 3--4\,MK temperatures, with peripheral $''$warm$''$ 2\,MK loops, and bright fan structures dominated by downflows emitting at lower temperatures (0.9\,MK) at the active region boundary. AR outflows mix in and around the downflows on the bright fans, and could be different structures, or part of the chromospheric-coronal mass cycle \citep{mcintosh_etal2012}. The outflows are more conspicuous and spatially extended at higher temperatures ($\sim$2\,MK), and have lower intensities in EUV images than the active region core and bright fans. We show example intensities and Doppler velocity maps from EIS observations made in support of the Hi-C 2.1 \citep[High resolution Coronal imager,][]{rachmeler_etal2019} sounding rocket flight of 29th May, 2018, in Fig. \ref{fig:fig2} \citep[see also][]{warren_etal2011a}. The Hi-C target region is AR 12712. Downflows on the fans (which are red in the Doppler velocity maps) and the outflows (which are blue) are marked with arrows. The velocity maps show the expected pattern of flows in solar active regions as described above. In particular, strong downflows (arrowed and labelled in the Figure) are seen on the bright fan structures to the Solar North East (NE) and South West (SW) in Fe IX 188.497\,\AA. These transition to downflows on the fans mixed with upflows in Fe XI 188.216\,\AA\, (labelled as mix) and Fe XII 195.119\,\AA\, (labelled as outflow) in the NE, and purely upflows in the SW (also labelled outflow). At the higher temperatures of Fe XIII 202.044\,\AA\, and above, we see only upflows. So the structures seen at different temperatures (Fe IX 188.497\,\AA\, and Fe XIII 202.044\,\AA, for example) are not the same.

Here we report observations of two apparent drivers of the outflows observed in AR 12712. We use high spatial resolution observations of the base of an outflow area from the Hi-C 2.1 flight, together with plasma composition measurements inferred from Hinode/EIS spectroscopic data. We describe the details of the observations in Section \ref{observations}. Plausible scenarios have been suggested for the formation of outflows into the fast solar wind in magnetic funnels in coronal holes \citep{tu_etal2005}. In contrast, it has been unclear where the outflows in active regions are originating from and what is driving them. One reason is their spatial extent in the corona, which is largest around 1.7\,MK, where they encompass areas containing many structures. Another reason is that at lower temperatures, where they appear more confined, the bright fan structures in active regions obscure the view, and make it difficult to understand which temperature represents the base of the outflows, and where they are ultimately emanating from. Furthermore, in low spatial resolution spectroscopic data, it is difficult to cleanly separate emission from different structures superimposing along the line-of-sight. It is also difficult to clearly determine the structures that are contributing to the emission. This is a particular problem for plasma composition measurements that generally require some kind of spatial averaging to improve the signal-to-noise ratio of the key diagnostic spectral lines observed in these dark areas. The high spatial resolution (250\,km) of the Hi-C narrow band images has allowed us to identify what features are contributing to the composition signature measured by EIS at lower spatial resolution ($>$2000\,km), and therefore to separate emission from those features and determine their relative contributions.
\section{Data Analysis and Results}

\subsection{Data sources and processing}

Hi-C 2.1 launched on 29th May 2018 and obtained 5 1/2 minutes of data from 18:56:21UT. We analyzed the complete Hi-C time-series of images, which were obtained at a fixed cadence of 4.4\,s. The Hi-C 2.1 bandpass is narrow (3\,\AA) and centered on 172\,\AA. The temperature response peaks at $\log$ (T/K) = 5.9 \citep{rachmeler_etal2019}, and is broader than that of the 171\,\AA\, filter of the Atmospheric Imaging Assembly \citep[AIA,][]{lemen_etal2012} on the Solar Dynamics Observatory \citep[SDO,][]{pesnell_etal2012}. The Hi-C 2.1 data we use have been calibrated to level 1.0 through removing the dark current and applying a flat-field from a master file obtained while the telescope was slewing to target during flight. Overscan bias pixels were removed and corrections for hot and dusty pixels applied. A total of 78 images were obtained. The instrument and performance of Hi-C during flight is discussed in a dedicated article \citep{rachmeler_etal2019}. We co-aligned each image to the nearest 171\,\AA\, image obtained by AIA. The AIA data we use were retrieved from the Joint Science Operations Center (JSOC) at Stanford and are also calibrated to level 1.0 following standard procedures \citep{boerner_etal2012}. The actual co-alignment was performed by first determining the Hi-C rotation angle and pointing offset from AIA using customized affine transformation software \citep{brooks_etal2012} and then optimizing the alignment using cross-correlation.  

\begin{figure*}[t!]
  \centerline{%
    \includegraphics[width=1.0\textwidth]{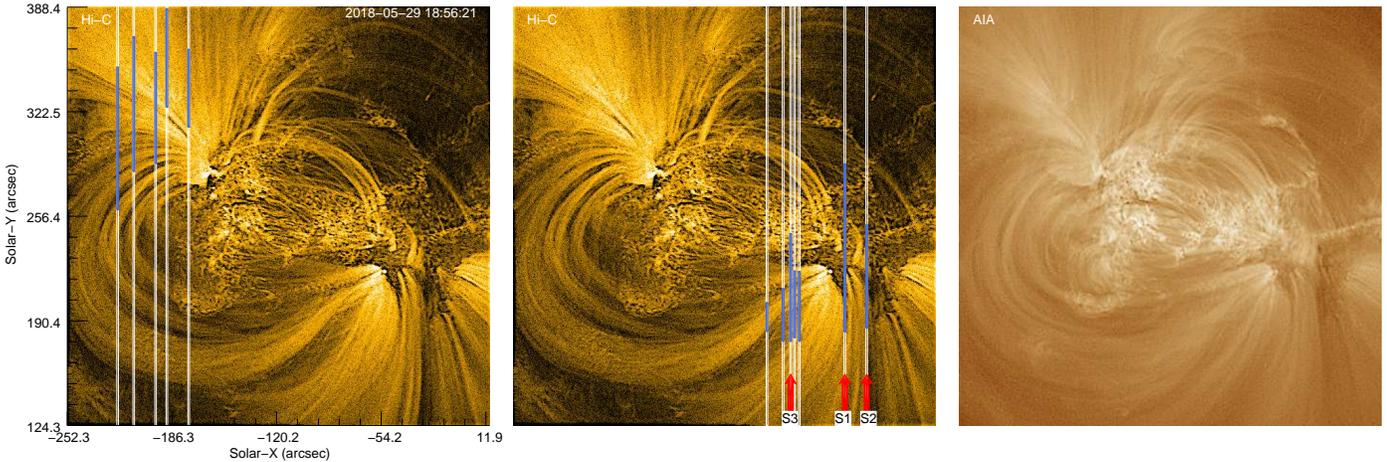}} %
  \caption{ Hi-C and AIA images of AR 12712 showing several EIS slit positions.
The images were taken at 18:56UT. These are narrow band images centered on 172\,\AA\, (Hi-C: left and middle panels) and 193\,\AA\, (AIA: right panel). They correspond to temperatures of 0.8 and 1.6\,MK, respectively). We sharpened all images for display, using a Gaussian filter. Several EIS slit positions are overlaid. The blue areas highlight regions of bulk upflows in the range of 5-30 \,km s$^{-1}$ in the NE (left panel) and SW (middle panel) along the slit. We exclude other lower velocity or isolated upflow patches along the slit positions. The slit positions marked S1 and S2 are the regions analyzed in Fig. \ref{fig:fig5} and Fig. \ref{fig:fig9}. The slit position marked S3 is also discussed in Section \ref{others}. Note that the EIS slit positions are not evenly distributed across the FOV. This is due to the orbital variation of the spectrum on the detector as a result of thermal instrumental effects. }
  \label{fig:fig3}
\end{figure*}

We used EIS on Hinode \citep{kosugi_etal2007} for the spectroscopic analysis of the outflows. EIS observes two wavelength bands from 171--211\,\AA\, and 245--291\,\AA\, with a spectral dispersion of 22.3\,m\AA\, per pixel. We processed the EIS data using standard calibration procedures available in SolarSoftware. These account for removal of hot, warm, and dusty pixels, CCD dark current, and strikes from cosmic rays. They also apply the radiometric calibration to convert the data from photon events to physical units (erg cm$^{-2}$  s$^{-1}$ sr$^{-1}$). To account for the evolving sensitivity of the instrument on orbit we applied the updated calibration of \citep{warren_etal2014}. 

Several EIS observations were planned in support of the Hi-C flight and we use two distinct datasets in this study. First, we make use of a wide (FOV) field-of-view (303$''$ $\times$ 384$''$) scan we obtained prior to the Hi-C launch window starting at 14:54:11UT. This scan uses the 2$''$ slit to cover the FOV in 3$''$ steps with 30\,s exposures. The second EIS dataset is from an observing program we designed specifically for observing during the rocket flight itself. It is a very coarse scan covering an FOV of 210$''$ $\times$ 512$''$ by moving the 1$''$ slit in 10$''$ steps. The exposure time for this program was 15\,s so it was able to complete a full scan of the AR during the Hi-C flight.
Both EIS datasets contain many diagnostic spectral lines covering a broad range of temperatures. 

The EIS raster scan data we use for context from prior to the rocket flight are taken stand alone, but we co-aligned the EIS coarse slit scan data obtained during the flight with the Hi-C images for detailed analysis. This was achieved by cross-correlating the EIS 195.119\,\AA\, intensities along the Y-direction of the slit with the intensities in the AIA 193\,\AA\, image taken closest in time to the EIS exposure, and locating the best match position. After co-alignment with AIA, the EIS data are automatically co-aligned to the pre-processed Hi-C images. A further refinement to the coalignment of EIS and AIA is to determine the EIS roll angle using the technique of \citep{pelouze_etal2019}.
We have not applied this correction here since our EIS measurements are made along the slit regardless of the roll angle, and the dynamic
activity we study in the Hi-C data are not detectable at the spatial and temporal resolution of EIS.

\subsection{Observations}
\label{observations}

In Fig. \ref{fig:fig3} we show a Hi-C image taken at 18:56:21UT with EIS slit positions from a rapid scan overlaid. These positions are coincident with upflows on the NE and SW (see Fig. \ref{fig:fig2}). We show the areas of upflow in blue in Fig. \ref{fig:fig3}. The velocities were determined using Fe XIII 202.044\,\AA\, (see Section \ref{velocity}). These blue regions highlight the boundaries of the bulk upflows in the range of 5-30\,km s$^{-1}$ on either side of the AR. We do not show other isolated upflow patches, or lower velocity areas, along the slit positions. Also, some patches within the upflows highlighted on the NE side (left panel) may include slight red-shifts on the fans. In fact, unfortunately, because of the superposition of foreground loop emission, or simply due to missing appropriate features, we could only use two slit positions for a detailed analysis. In all the positions on the NE upflow, for example, the slit is crossing the bright fans. Fig. \ref{fig:fig2} shows that these bright fans are strongly red-shifted at lower temperature. So even when upflows are detected near 1.7\,MK in Fe XIII 202.044\,\AA, the emission from lower temperatures is coming from different structures that show downflows. It is possible that there are outflows at these lower temperatures along the line-of-sight, but the bright fans block our view of what is below. So we cannot isolate any outflow component at low temperatures, and an emission measure analysis becomes inappropriate.

The situation is more promising on the SW side (middle panel). There the bright structures are more closely aligned with the large loops connecting the leading negative polarity to the trailing positive polarity of the AR. They also do not fan out so dramatically to obscure the view. So several EIS slit positions where upflows are detected cross the fans from the south and offer an unobstructed view of the root of the outflows. These are the key positions for our analysis.

\begin{figure*}[t!]
  \centerline{%
    \includegraphics[width=1.0\textwidth]{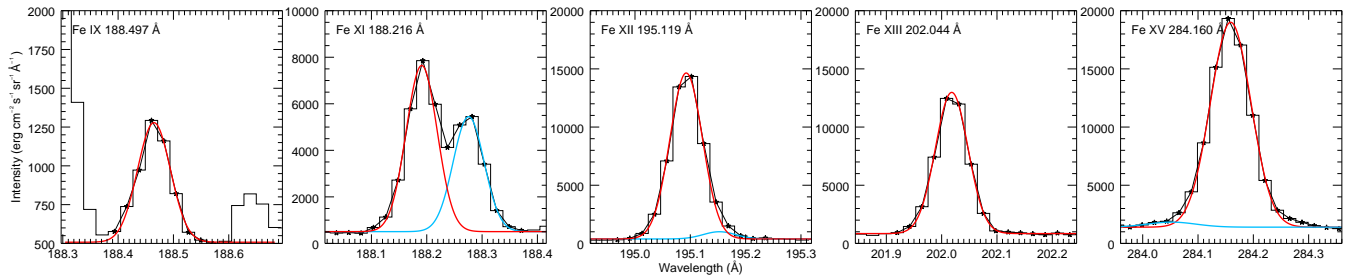}} %
  \caption{Example line profiles in the outflow region.
Spectral line profiles for the averaged bulk outflow (white box in Fig. \ref{fig:fig5}). The spectral lines are indicated in the legends and are the same ones we used to construct the intensity images and Doppler velocity maps in Fig. \ref{fig:fig2}. They cover the temperature range of 0.87 to 2\,MK. The black histograms and stars show the EIS data. The red lines show the Gaussian fits to the lines. The light blue lines show Gaussian fits to minor blending lines. }
  \label{fig:fig4}
\end{figure*}

\begin{deluxetable*}{cccccccccc}
\tabletypesize{\scriptsize}
\tablewidth{0pt}
\tablecaption{Spectral lines used for analysis. \label{table}}
\tablehead{
\multicolumn{1}{c}{Element} &
\multicolumn{1}{c}{Ion} &
\multicolumn{1}{c}{$\lambda$/\AA}&
\multicolumn{1}{c}{T$_f$/MK}&
\multicolumn{1}{c}{I$_a$}&
\multicolumn{1}{c}{$\sigma_a$}&
\multicolumn{1}{c}{I$_b$}&
\multicolumn{1}{c}{$\sigma_b$}&
\multicolumn{1}{c}{I$_p$}&
\multicolumn{1}{c}{$\sigma_p$}
}
\startdata
Fe[7.9] & IX   & 188.497 & 0.9 &     62.4 &     14.3 &   2845.8 &    672.1 &   1136.9 &    349.6\\
Fe &      IX   & 197.862 & 0.9 &     40.5 &      9.2 &   1410.1 &    336.8 &   1089.6 &    273.2\\
Fe &      X    & 184.536 & 1.1 &    371.7 &     83.1 &  17906.3 &   4038.7 &   5222.1 &   1453.4\\
Fe &      XI   & 188.216 & 1.4 &    503.0 &    110.9 &  20397.9 &   4508.8 &  10286.2 &   2304.9\\
Fe &      XII  & 195.119 & 1.6 &    930.5 &    204.9 &  37523.2 &   8268.8 &  19238.0 &   4259.0\\
Fe &      XIII & 202.044 & 1.8 &    596.3 &    131.8 &  28234.3 &   6258.4 &   8143.3 &   1947.7\\
Fe &      XIII & 203.826 & 1.8 &    519.8 &    116.2 &  16201.0 &   3781.1 &  15538.3 &   3644.0\\
Fe &      XIV  & 264.787 & 2.0 &    688.4 &    151.6 &  34470.2 &   7596.7 &   8795.2 &   1986.3\\
Fe &      XIV  & 270.519 & 2.0 &    407.7 &     89.8 &  14001.8 &   3095.4 &  11101.0 &   2461.1\\
Fe &      XV   & 284.160 & 2.2 &   4312.0 &    948.9 & 191643.7 &  42187.5 &  71489.0 &  15796.9\\
Fe &      XVI  & 262.984 & 2.8 &    300.1 &     66.3 &  12430.9 &   2762.0 &   6255.6 &   1429.6\\
Ca[6.1] & XIV  & 193.874 & 3.5 &     21.9 &      5.0 &   1216.5 &    280.4 &    310.4 &    107.9\\
Si[8.2] & X    & 258.375 & 1.4 &    440.5 &      8.2 &  15298.4 &   3402.6 &  11612.6 &   2603.2\\
S[10.4] & X    & 264.233 & 1.5 &    144.3 &      4.7 &   3905.2 &    904.6 &   5209.0 &   1180.4
\enddata
\tablenotetext{*}{$\lambda$ is the wavelength. T$_f$ is the formation temperature.
I$_a$ is the intensity of the averaged bulk outflow. I$_b$ is the total intensity in the background/foreground emission.
I$_p$ is the total intensity in the active region plage. The uncertainties in the intensities are denoted by $\sigma$ and
include the calibration error added in quadrature. The first ionisation potentials for each element are given in brackets
in eV. Units are erg cm$^{-2}$ s$^{-1}$ sr$^{-1}$. 
}

\end{deluxetable*}

The Hi-C bandpass is dominated by emission from the strong Fe IX 171.073\,\AA\, resonance transition formed at $\sim$0.8\,MK. So images such as those in Fig. \ref{fig:fig3} clearly show emission from fan structures and bright plage. The AIA 193\,\AA\, filter is dominated by emission formed at 1.6\,MK\, and so the image in Fig. \ref{fig:fig3} (right panel) provides context at higher temperatures. The EIS observations in Fig. \ref{fig:fig2} show intensity images of the target active region prior to the Hi-C launch window and provide further diagnostic information from the extra wavelength dimension (used to construct velocity maps). To compliment the intensity and velocity maps in Fig. \ref{fig:fig2}, we show example spectral line profiles from one of the outflow areas in Fig. \ref{fig:fig4}. These are mean profiles from the white boxed area shown in the top panel of Fig. \ref{fig:fig5}. 
We fit the spectra for all the lines we analyzed mostly using single Gaussian functions, but also took account of specific known blends using multiple Gaussian fitting.
There are no strong asymmetries in the profiles indicating that a single Gaussian is a good fit in the areas of outflow we analyze here, but the profiles are not representative of all outflows in every active region. Other regions show a high-speed blue wing component and the asymmetry increases as a function of temperature, see e.g. Figure 2 in \citep{brooks&warren_2012}. In other regions, a down flow component may be detectable. The strength of any red or blue shifts also depends on the orientation to the observer's line of sight \citep{mcintosh_etal2012}.

\subsection{Velocity measurements}
\label{velocity}
The EIS instrument does not have an absolutely calibrated wavelength scale. Therefore, we quote only relative Doppler velocities in this article. These were calibrated as follows. First, an artificial neural network model was applied to the data to remove the orbital drift of the spectrum on the detector \citep{kamio_etal2010}. This model uses satellite housekeeping temperature information to correct the drift and is expected to be accurate to $\sim$4.5\,km s$^{-1}$. Since the spectrum is moving on the CCD, and real plasma motions on the Sun are changing the positions of the spectral lines, a reference wavelength is needed to convert from the measured line centroids to Doppler velocities. The neural network model assumes that the Doppler shift of the strong Fe XII 195.119\,\AA\, line is zero when averaged over the entire mission dataset. There is evidence from absolutely calibrated spectra, however, that in fact the corona is slightly blue-shifted at the formation temperature of Fe XII \citep{peter&judge_1999}. Therefore, we determined reference wavelength sfor the lines we use by averaging the fitted line centroids in the upper 100 pixels of the CCD as far away from the active region as possible within the EIS FOV. When quoting results, later in the discussion of Fig. \ref{fig:fig5}, we also apply a correction to the velocities to account for the calibrated on-disk coronal blue-shift \citep{peter&judge_1999}. For the Fe XIII 202.044\,\AA\, line this is at least 4.5 \,km s$^{-1}$. In a final step, we removed a residual orbital drift that remained after the standard correction. This is discussed briefly in the Appendix.

The biggest uncertainty associated with this technique is the choice of the reference wavelength. Given the EIS FOV and the extent of the active region, the upper portion of the CCD could not realistically be classified as truly quiet Sun. Transition region emission from the strong Si VII 275.368Å line in this region is about a factor of 2.4 higher than measured in the quiet Sun \citep{brooks_etal2009}. In this work, however, we are not overly concerned with the absolute values of the velocities. We only use the relative velocity maps to identify regions of upflow. Typical Doppler shifts in coronal lines such as Fe XII 195.119\,\AA\, in active regions are only 5-10\,km s$^{-1}$ \citep{delzanna_2008}, whereas the outflows typically show larger bulk velocities of 10-40\,km s$^{-1}$, with wings in the line profiles extending to much higher velocities \citep{brooks&warren_2011,brooks&warren_2012}.

\subsection{Composition measurements and emission measure distributions}
We compute the plasma composition, commonly referred to as the FIP bias, by first determining the plasma electron density and temperature structure from the equation
\begin{equation}
I_{ij} = A(Z) \int \phi (T) G_{ij} (T,n) dT
\end{equation}
where $I_{ij}$ is the line intensity for a transition from level $j$ to $i$ within a particular ion, $A(Z)$ is the elemental abundance of species $Z$,  $\phi (T)$ is the differential emission measure (DEM) as a function of temperature, $T$, and is defined as $\phi (T) = n^2 ds/dT$, where $n$ denotes the electron density and $ds$ is the path length along the line of sight. Throughout our analysis we refer to the emission measure (EM) distribution which we define as $\phi (T) dT$. $G(T,n)$ is the contribution function that contains all of the necessary atomic physics coefficients (spontaneous radiative decay, upper level population, ion fraction etc). This equation definition makes several simplifying assumptions that have been discussed in great detail in the literature \citep{craig&brown_1976,lang_etal1990,judge_etal1997}. In particular, since $G(T,n)$ is dependent on both $T$ and $n$, we estimate the electron density in order to compute this function to the highest accuracy for all the spectral lines used in the EM analysis. 

Here we use the Fe XIII 202.044/203.826 diagnostic ratio to measure the plasma density. This ratio varies by a factor of 120 in the range of $\log$ (n/cm$^{-3}$) = 7--10. This density is then used to compute the contribution functions using the CHIANTI v8.0 database \citep{delzanna_etal2015}, assuming photospheric abundances \citep{grevesse_etal2007} and the CHIANTI ionization fractions \citep{dere_etal2009}. We use spectral lines from Fe IX-XVI together with Ca XIV 193.874\,\AA\, and Si X 258.375\,\AA\, to determine the temperature distribution of the feature of interest. The specific lines used are listed with their formation temperatures in quiet Sun conditions in Table \ref{table}. With the exception of the Fe XIII 202.044\,\AA\, and Fe XIII 203.826\,\AA\, density sensitive lines, the contribution functions for all the other spectral lines are mostly insensitive to density; the $G(T,n)$ peak magnitudes vary less than 25\% for the typical density range of the outflows, log n = 8.4-9.0 \citep{brooks&warren_2011}. 

\begin{figure*}[t!]
  \centerline{%
    \includegraphics[width=1.0\textwidth]{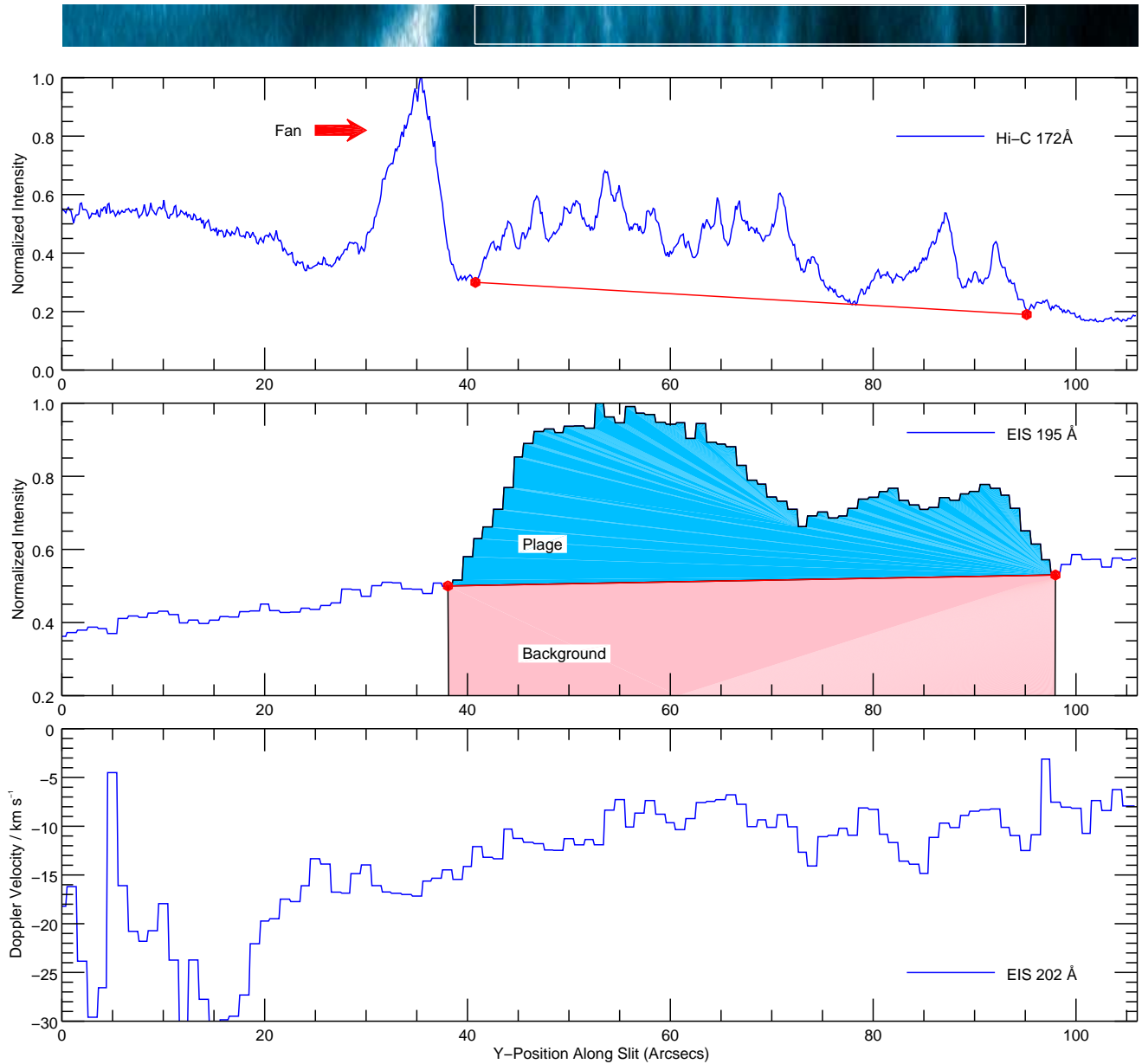}} %
  \caption{Hi-C and EIS observations of the outflow area detected by the EIS slit.
Top row: rotated Hi-C image of the outflow area detected at one EIS slit position (labeled S1 in Fig. \ref{fig:fig3}). The aspect ratio of the image is reduced from the original in order to better reveal the features. The Hi-C plate scale is ~8 times better than EIS, so a single EIS pixel in the E-W direction corresponds to at least 8 Hi-C pixels (the actual value depends on the ratio of the EIS and Hi-C point-spread-functions). The white box shows the plage region of interest. Second row: normalized Hi-C intensity along the slit averaged in the solar E-W direction.  The bright fan footpoint is highlighted by an arrow. The red dots show the locations chosen for background/foreground subtraction and the red line shows the polynomial fit between these positions. Third row: normalized EIS 195.119\,\AA\, intensity along the slit. The plage and background/foreground components are highlighted in sky blue and pink, respectively. Bottom row: Doppler velocity measured in the EIS 202.044\,\AA\, spectral line, showing that this whole region is within the blue-shifted outflow. }
  \label{fig:fig5}
\end{figure*}

\begin{figure*}[t!]
  \centerline{%
    \includegraphics[width=1.0\textwidth]{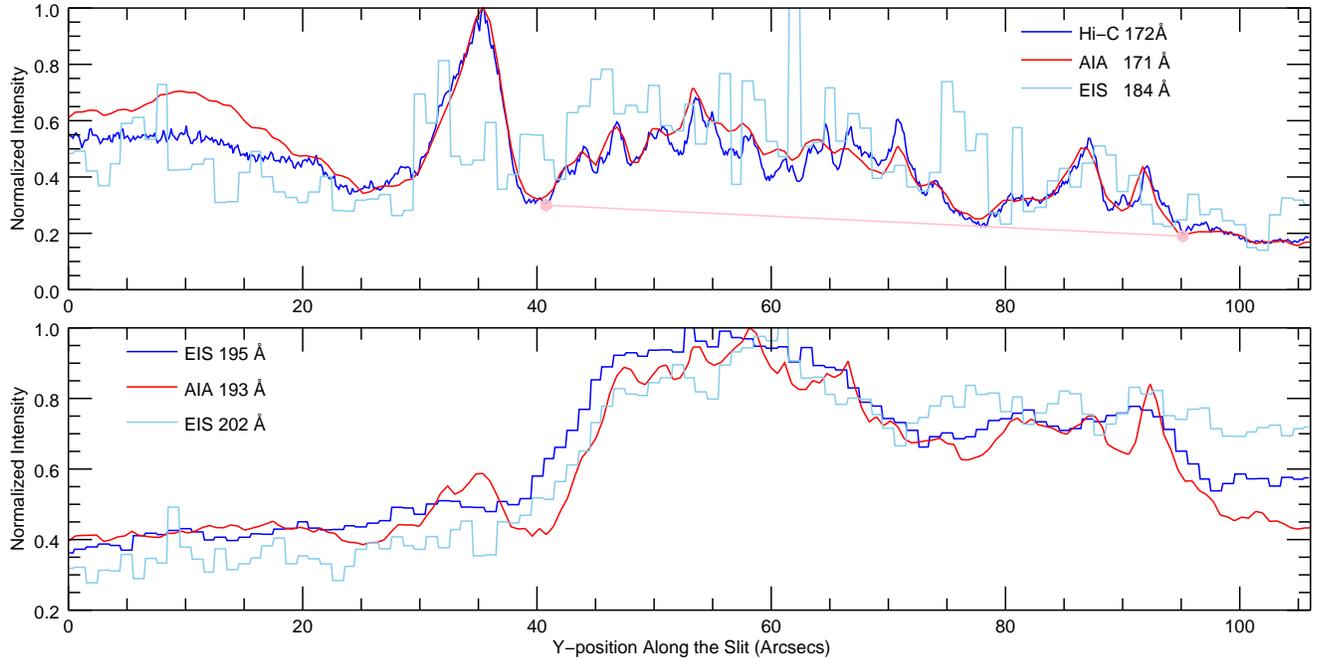}} %
  \caption{Hi-C, EIS, and AIA observations of one slit position.
A comparison of the Hi-C, EIS, and AIA data for the same slit position as shown in Fig. \ref{fig:fig5}. Top row: normalized Hi-C intensity along the slit averaged in the solar E-W direction (blue) with the normalized AIA 171\,\AA\, intensity (red) and EIS Fe X 184.536\,\AA\, intensity (sky blue) overlaid. The pink dots show the locations chosen for background/foreground subtraction and the pink line shows the polynomial fit between these positions. Bottom row: normalized EIS Fe XII 195.119\,\AA\, intensity along the slit (blue) with the normalized AIA 193\,\AA\, intensity (red) and EIS Fe XIII 202.044\,\AA\, intensity (sky blue) overlaid. The Hi-C and AIA 171\,\AA\, data are highly correlated and it is clear that the dots are a good representation of the boundary of the plage in both the Hi-C and AIA data. Most of the dynamic features observed by Hi-C are also detected by AIA, though their intensities and widths are better constrained by the higher spatial resolution Hi-C data. The EIS Fe X 184.536\,\AA\, intensities are not strongly correlated with the Hi-C intensities, and the dynamic features and plage boundary are already difficult to locate in the lower spatial resolution data. At higher temperatures the EIS Fe XII 195.119\,\AA, Fe XIII 202.044\,\AA, and AIA 193\,\AA\, intensities are well correlated. }
  \label{fig:fig6}
\end{figure*}

\begin{figure*}[t!]
  \centerline{%
    \includegraphics[width=0.32\textwidth]{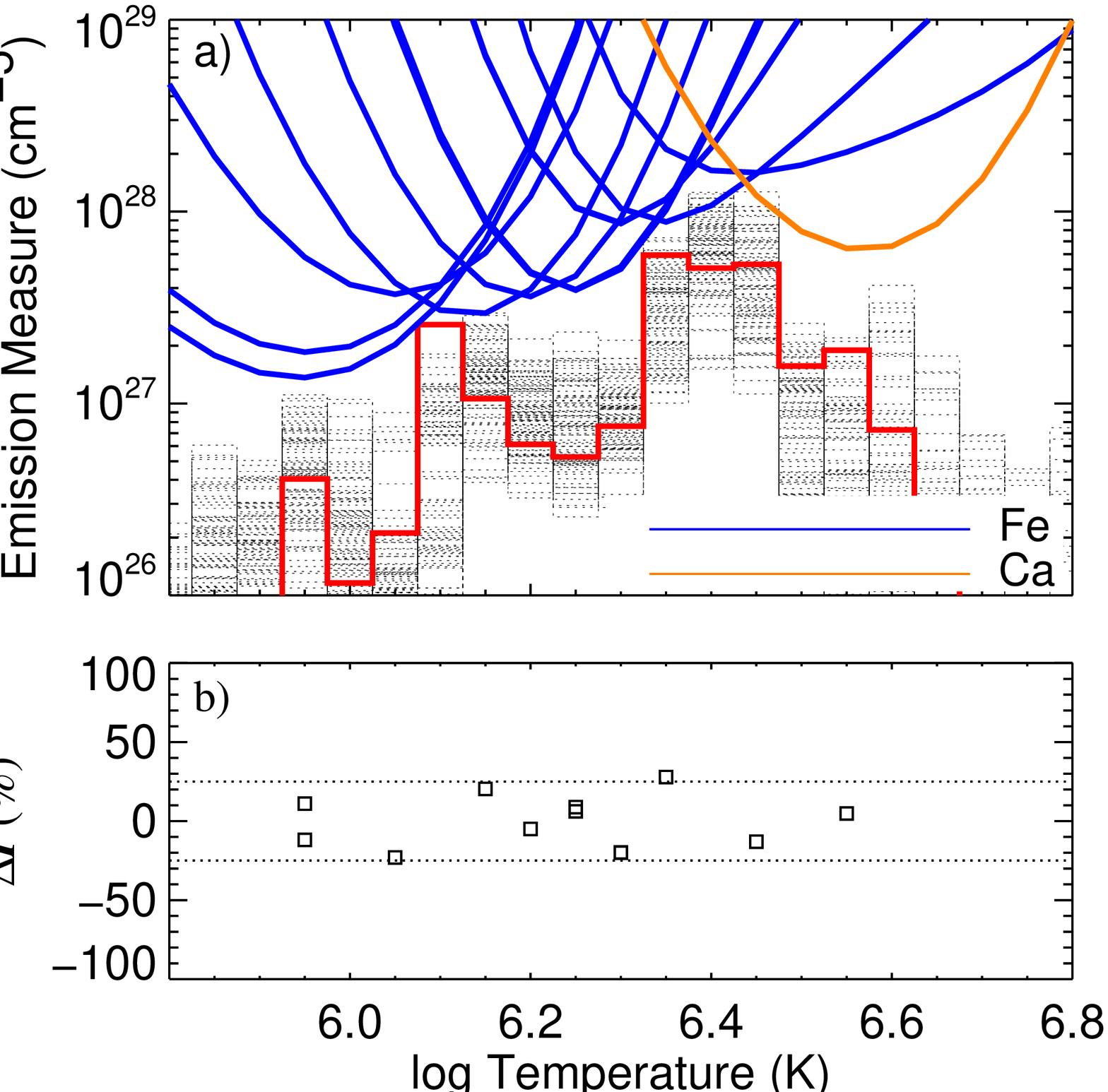} %
    \includegraphics[width=0.32\textwidth]{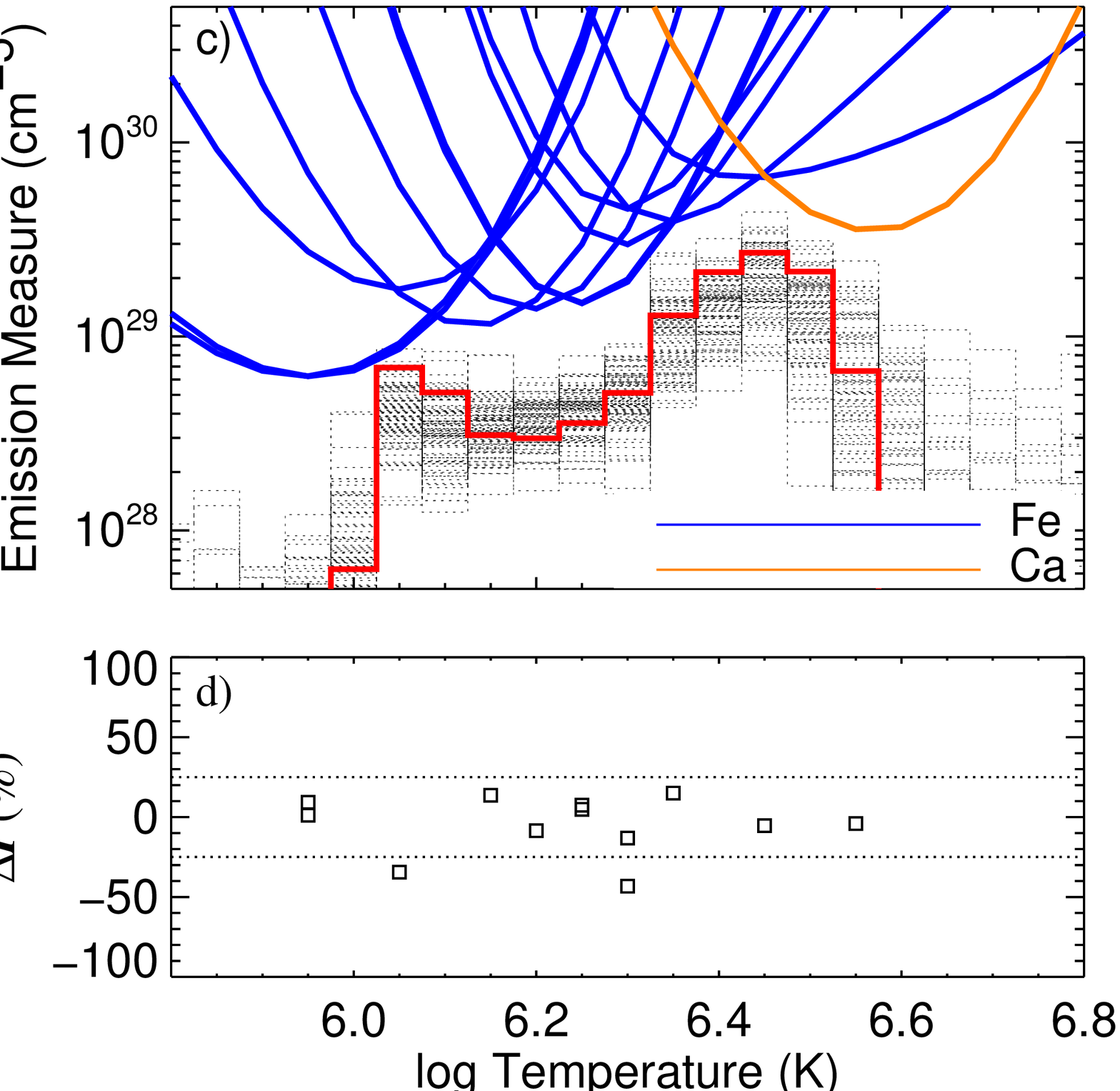} %
    \includegraphics[width=0.32\textwidth]{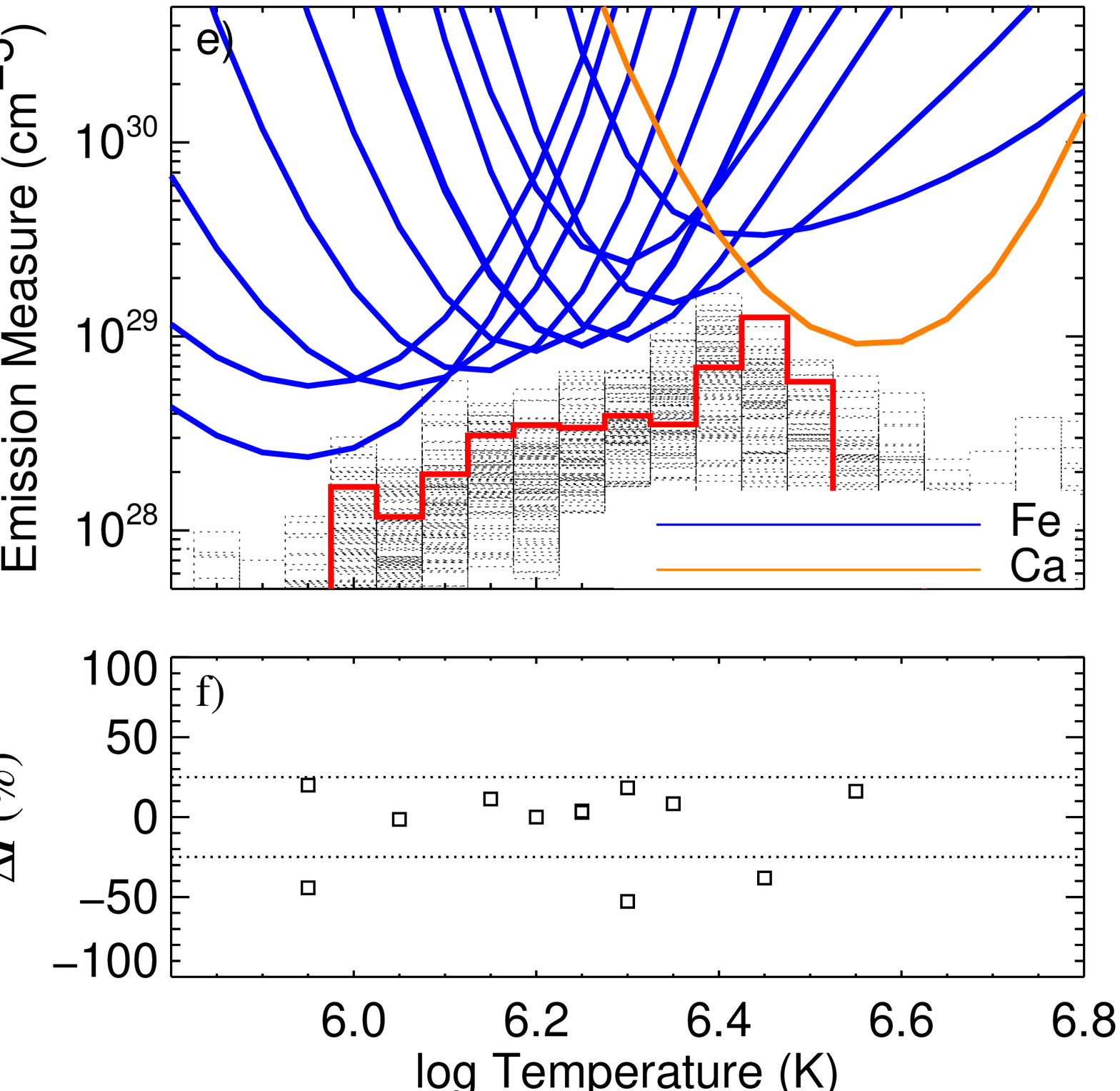}} %
  \caption{Outflow component temperature distributions.
Emission Measure (EM) distributions as a function of temperature for the averaged bulk outflow (a), the total intensity in the background/foreground emission (c), and the total intensity in the active region plage (e). These features are described in more detail in the text associated with Fig. \ref{fig:fig5}. The red curves show the best-fit EM distributions, and the dotted grey lines show the Monte Carlo simulations. The solid blue and orange lines show EM loci curves for the Fe and Ca spectral lines used in the analysis, and indicate upper limits to the EM distributions assuming zero intensity. The lower panels (b, d, and f) show the differences between the observed and computed line intensities expressed as percentages of the measured intensity. The dashed lines show the boundary levels where differences are below 25\%. The electron densities for the three regions (a, c, e) are $\log$ (n/cm$^{-3})$ = 8.9, 8.7, and 9.3, respectively. The corresponding FIP bias values are 1.8, 2.5, and 1.0. }
  \label{fig:fig7}
\end{figure*}

We compute the emission measure using the Markov-Chain Monte Carlo algorithm available in the PINTofALE software package \citep{kashyap&drake_1998,kashyap&drake_2000}. This method reconstructs the temperature distribution by estimating the amount of emission measure needed to reproduce all of the observed line intensities. In our analysis, we compute 100 potential realizations of the emission measure distribution from the Monte Carlo simulations and we use the solution that best fits the data. Since Fe, Ca, and Si are all low FIP elements, we expect the temperature distributions derived from spectral lines of these elements to be similar. Initially, we only use the Fe lines for the actual derivation - to minimize the influence of the choice of elemental abundances - with Ca XIV acting as a high temperature constraint. We then make an adjustment to match the Si X 258.375\,\AA\, intensity, and the final temperature distribution is used to simulate the expected intensity of the S X 264.233\,\AA\, line.

The $G(T,n)$ functions for Si X 258.375\,\AA\, and S X 264.223\,\AA\, are very similar. \citet{brooks&warren_2011} show them in their Figure 1, and discuss the range of validity
of the ratio in terms of densities and temperatures. To achieve the highest accuracy, the density should be measured and the G(T,n) functions convolved
with the emision measure distribution, as we do here. The low- and high-FIP groups of elements are usually defined as hvaing a FIP below or above 10\,eV. With no enhancement of low-FIP elements, we expect the density and temperature distribution that reproduces Si X to be valid for S X. If
the low FIP elements are enhanced, however, the prediction for S X 264.223\,\AA\, will be too large because it is a high FIP element. The ratio of the
predicted to observed intensity of S X 264.223\,\AA\, then gives the FIP bias. Note that with a FIP of 8.2\,eV Si has the highest FIPs of the
low-FIP group of elements, and with a FIP of 10.4\,eV S is very close to the boundary between groups. So these elements are not necessarily the best ones to use for detecting a strong 
FIP effect. In particular, S sometimes shows behavior that could be described as intermediate between low- and high-FIP elements \citep{reames_2018}. In theoretical models,
this depends on whether the magnetic field is open or closed, and therefore whether Alfven waves can achieve resonance \citep{laming_2015}. 
We therefore stress here that the FIP bias we are measuring is, strictly speaking, the ratio of the Si and S coronal abundances. Note, however, that
the EIS composiiton measurements made using this ratio have been quite successful in capturing the expected trends of the FIP effect. For example,
a photospheric composition is detected in polar coronal holes \citep{brooks&warren_2011}, and an enhanced composition is detected in bright active
region loops \citep[see e.g. ][]{doschek&warren_2019}. Ideally the results would be checked against measurements made with other elements, but useful spectral lines from
other high FIP elements are mostly emitted at higher temperatures \citep{feldman_etal2009b}.

Our technique has been well tested for robustness in previous studies \citep{brooks&warren_2011,baker_etal2015,brooks_etal2015}, and specifically
to assess the impact of potential cross-calibration problems between the short- and long-wavelength detectors, a significant difference in fractionation
behavior between Fe and Si, unknown problems with the atomic data, and to determine the uncertainties in the computed FIP bias measurements
\citep{brooks_etal2017}. 
The conclusion was that the method works well even if the short- to long-wavelength calibration, or Si/Fe fractionation, is in error because we are only using Fe lines on the short-wavelength detector to determine the shape of the DEM, not the magnitude. The magnitude is determined by the intensity of Si X 258.375\,\AA. Furthermore, the most relevant part of the DEM (near 1.5\,MK) is dominated by emission from lines of Fe XI, Fe XII, and Fe XIII, which are relatively close in wavelength. The effect of the test on atomic data uncertainties is to modify the DEM and
produce a dispersion in FIP bias values of $\pm$0.3. This is the estimated uncertainty.
Since these were generic experiments on how the method handles input intensities and atomic data they are applicable here. Of course we
also assume that the simplifying assumptions of the DEM method are valid. If that were not the case, for example if the plasma is not
in ionisation equilibrium, then there could be significant systematic errors in the atomic data. We expect that these would be revealed as systematic
deviations between observed and calculated intensities.

In Fig. \ref{fig:fig5} we show the analysis of one of the EIS slit positions. The Figure shows the structures observed by Hi-C in the upper transition region (formation temperature of the 172\,\AA\, filter), and corresponding features seen in the 195.119\,\AA\, spectral line in the low corona by EIS. In this case, the EIS slit (labeled S1 in Fig. \ref{fig:fig3}) is fortuitously positioned between the fans. The lower part of the slit does glance the footpoint of a bright fan loop around pixel positions 30--40 (highlighted with an arrow in Fig. \ref{fig:fig5}). The upper part of the slit, however,  passes across an extended area of active region plage, or moss-like emission around pixels 40--95. Moss is usually defined as the footpoints of high temperature loops \citep{berger_etal1999}. Analysis of the high temperature emission in this AR suggests that loops seen in Fe XVIII are not connecting to the regions we identify as outflow \citep{warren_etal2020}. So we refer to these moss-like areas as plage here. We see considerable structure in the plage in Hi-C, but it appears fairly homogenous at the higher temperature of Fe XII 195.119\,\AA. This is primarily due to the relatively lower spatial resolution of EIS (see Fig. \ref{fig:fig6}).

The Hi-C data enable us to understand what EIS is measuring, and exclude for example the region around the fan footpoint that is not detectable in Fe XII 195.119\,\AA\, but gives a contribution to the 172\,\AA\, emission from down flowing plasma. Furthermore, the Hi-C data also play another important role in defining where to extract the background/foreground emission from the structures of interest. Until now, all outflow measurements have been made without treating this background/foreground emission when determining the temperature distribution and using it to infer the plasma composition. Yet this has proven to be critical in the analysis of the temperature distributions of coronal loops \citep{klimchuk&porter_1995,delzanna&mason_2003}. Since the outflow has expanded above and around the plage and fan loops, emission from the outflow has itself become part of the background measured by EIS. So it is not clear how to isolate this contribution. The Hi-C data, however, clearly show where the plage is delineated, so the plage component can be extracted and the outflow component is what remains. We show the positions of the boundaries of the plage as red dots in Fig \ref{fig:fig5}. We fit a polynomial to these background positions and extract the plage emission above the linear fit (sky blue region in Fig \ref{fig:fig5}). This is a simple approximation, but is similar to the analysis that has been done for coronal loops \citep{aschwanden&nightingale_2005, warren_etal2008a}. The uncertainties will likely be larger here since the fit is made over a greater distance, but the method is supported by previous observations that show that the background emission increases approximately linearly from the periphery to the core of ARs \citep{delzanna_2013b} with a gradient that is instrument dependent. The background/outflow component (pink region in Fig \ref{fig:fig5}) is the emission below the linear fit. We then use the approximate locations of the background positions to extract the intensities of the two components for all the other EIS spectral lines used in the analysis. The intensities are totaled between the red dots. This procedure is done automatically to reduce any bias introduced by visual selection. Future observations with high spatial resolution at all wavelengths will help to confirm these measurements.

We show the resulting emission measure distributions for three examples in Fig. \ref{fig:fig7}. These distributions are for the outflow region analyzed in Fig. \ref{fig:fig5}. That is, the mean outflow, the background/foreground outflow, and the plage region. The electron densities measured using the Fe XIII are $\log$ (n/cm$^{-3})$ = 8.9, 8.7, and 9.3 for these regions, respectively, with an uncertainty on the measured ratio of $\sim$30\% (0.12-0.16 dex) due to the instrument photometric calibration. In most cases the differences between the observed and calculated spectral line intensities are within $\sim$25\%: 11/12 of the lines in the mean outflow, 10/12 in the background/foreground outflow, and 9/12 in the plage region. This indicates that the emission measure distributions are well constrained. Most of the few discrepant lines (including the worst - Fe XIV 270.519\,\AA\, - which is 50\% out in the plage region) emit at temperatures above 2\,MK. This is considerably higher than the temperature of the Hi-C 172\,\AA\, filter where detailed features are observed in the plage and therefore may be emission coming from unconnected structures. It is also far from the temperature where the FIP bias is measured.

Our technique assumes that Si and Fe fractionate in a similar way due to the FIP effect because they are both low FIP elements. It appears, however, that that might not always be the case \citep{heidrich_etal2018}, and in making our composition measurements we found some evidence that Si is fractionated more than Fe. For example, in the worst case in the plage region, the difference between the observed and calculated Si X 258.375\,\AA\, intensity is $\sim$33\% before the Fe and Si emission measure distributions are matched. This is larger than  the intensity calibration error \citep{lang_etal2006} so the difference could be real. It is difficult to definitively pin down the reason for this difference, however. It could be that it just reflects a difference in the degradation of the intensity calibration between the short- and long-wavelength detectors. It is also possible that a small error in the atomic data is showing up here in the most marginal case. A fractionation-split of this magnitude is also much smaller than the large difference in the composition signature we measure in the plage and outflow so does not affect this result. It does, however, remind us to note that the FIP bias we are measuring is, strictly speaking, the ratio of the Si and S coronal abundances.
\begin{figure*}[t!]
  \centerline{%
    \includegraphics[width=1.0\textwidth]{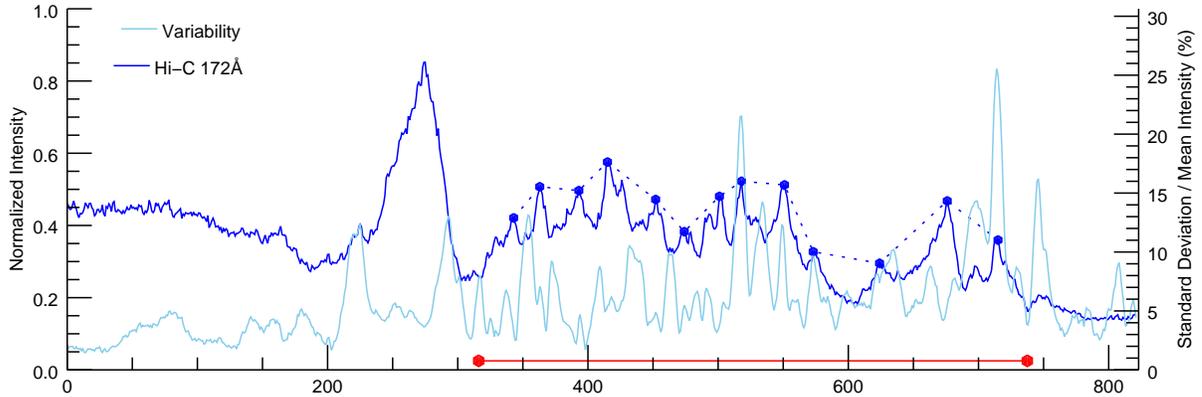}} %
  \caption{Time variability of features observed in the plage by Hi-C.
Normalized Hi-C intensity (dark blue) for the same slit position as in Fig. \ref{fig:fig5}. We highlight the bright structures in the plage (discussed in the text) with the blue dots. We measured the widths, lifetimes, and intensity variations of these features and they show an average intensity variation of 30\%. The sky blue line shows the ratio of the standard deviation of the intensities (in time) to the mean intensity (in time) expressed as a percentage. This measure shows a lower level of variability but is 15-25\% at several locations even over the short time interval of the rocket flight, which is larger than that seen at lower spatial resolution in EIS observations; less than 15\% over many hours \citep{brooks&warren_2009}. The red dots show the locations of the boundaries of the plage for reference. These locations were chosen for the background/foreground subtraction. }
  \label{fig:fig8}
\end{figure*}

\begin{figure*}[t!]
  \centerline{%
    \includegraphics[width=1.0\textwidth]{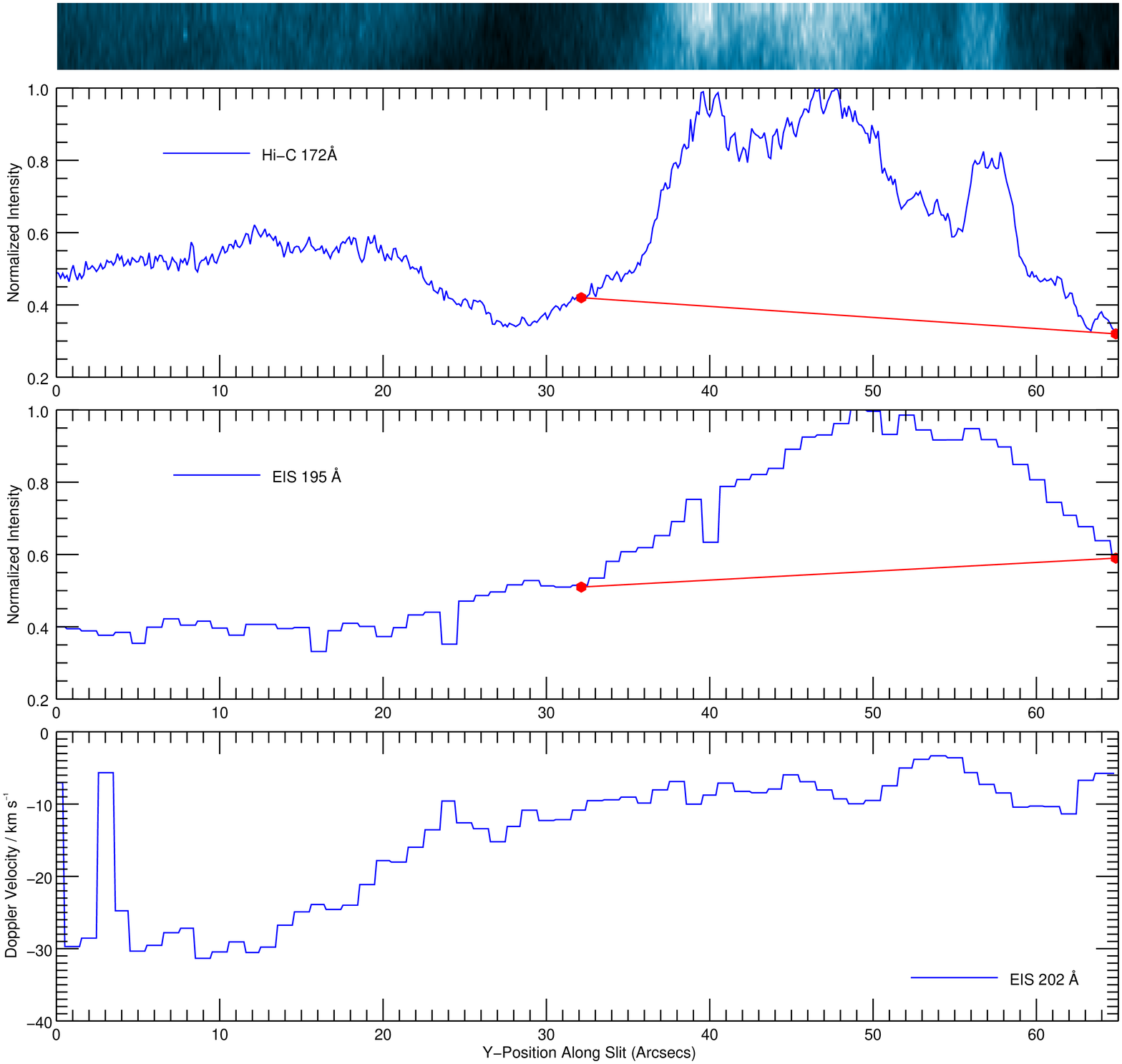}} %
  \caption{Hi-C and EIS observations of an outflow area detected by the EIS slit.
Same as Fig. \ref{fig:fig5} but for a different position of the EIS slit. Top row: rotated Hi-C image of the outflow area detected at the EIS slit position S2 (extreme right position in the left hand panel of Fig. \ref{fig:fig3}). The aspect ratio of the image is reduced from the original size in order to better reveal the features. Second row: normalized Hi-C intensity along the slit averaged in the solar E-W direction. Third row: normalized EIS 195.119\,\AA\, intensity along the slit. The red dots show the locations chosen for background/foreground subtraction and the red line shows the polynomial fit between these positions. Bottom row: Doppler velocity measured in the EIS 202.044\,\AA\, spectral line, showing that this whole region is within the blue-shifted outflow. }
  \label{fig:fig9}
\end{figure*}

Our measurements show that the FIP bias in the outflow region of interest (white box in Fig. \ref{fig:fig5}) is 1.8. That is, the outflows show a coronal composition, consistent with previous measurements in similar active regions and the expectation for the slow solar wind. The emission from the expanded background/foreground component of the outflow shows a higher FIP bias of $\sim$2.5. This is a new result because this component of the outflow has been separated from the plage emission.

The composition in the plage region is different. The FIP bias is $\sim$1.0, which indicates photospheric abundances. Plage regions are known to be sites of dynamic activity from chromospheric jets \citep{depontieu_etal2007a,depontieu_etal2009}. The moss emission, however, generally shows lower variability when observed with lower spatial resolution in the corona \citep{brooks&warren_2009}, except around the edges where the magnetic field is changing, or at the footpoints of the hottest loops \citep{testa_etal2013}. We might therefore expect an increased range of variability at the temperature and spatial resolution of Hi-C, and we do detect some evidence of this. The bright structures seen in the Hi-C image and intensity plot of Fig. \ref{fig:fig5} show an average intensity variation of 30\% (range 9--86\%) on a mean timescale of 167s (range 65--243s). These dynamic properties were measured using only the low jitter images. We also show the variability of the emission in Fig. \ref{fig:fig8}. Without simultaneous lower temperature observations we cannot connect this activity to specific chromospheric features such as type II spicules or dynamic fibrils, but the lifetimes and spatial scales (widths of 232--692\,km with an average of 503\,km) are comparable to the lifetimes and widths of these chromospheric jets \citep{depontieu_etal2007b,pereira_etal2012}.

It is worth noting that one of our experiments showed that the uncertainty in the FIP bias could be greater than 30\% if the errors in the atomic data are larger
than $\sim$40\%. We do not expect the atomic data errors to be this large, but in any case a 30\% error is within the intensity calibration uncertainty for the Si X/S X ratio, and is much smaller than the factor of 2.5 difference between
the FIP bias measurements in the plage region and background subtracted outflow. Our key result is that a photospheric composition is detected in the plage with a
FIP bias of $\sim$1.0 The value for the bulk outflow is an average and is not uncorrelated with the other two measurements.

\citet{delzanna&mason_2018} give a
thorough review of the current status of elemental abundance studies in the literature for different solar features. These measurements have been derived from a
variety of methods, and they do not all agree. At least some of these discrepancies are due to differences in the emission measure analysis techniques and the assumptions associated with that.

\subsection{Analysis of other EIS slit positions}
\label{others}

We also examined the two other EIS slit positions that cross into the plage (labeled S2 and S3 in Fig. \ref{fig:fig3}), using the same methodology as for S1. The slit position to the extreme right (S2) showed a similar coronal composition for the outflow region (FIP bias $\sim$2), though greater variability in the S X 264.233\,\AA\, intensity distributions made the measurements of the separate components very sensitive to the background subtraction. Ultimately we were not able to obtain a satisfactory solution for the two components simultaneously, and could not confirm the presence of photospheric composition plasma in the plage for this case. Fig. \ref{fig:fig9} is similar to Fig. \ref{fig:fig5} and shows this example but without the separated components.

\begin{figure*}[t!]
  \centerline{%
    \includegraphics[width=1.0\textwidth]{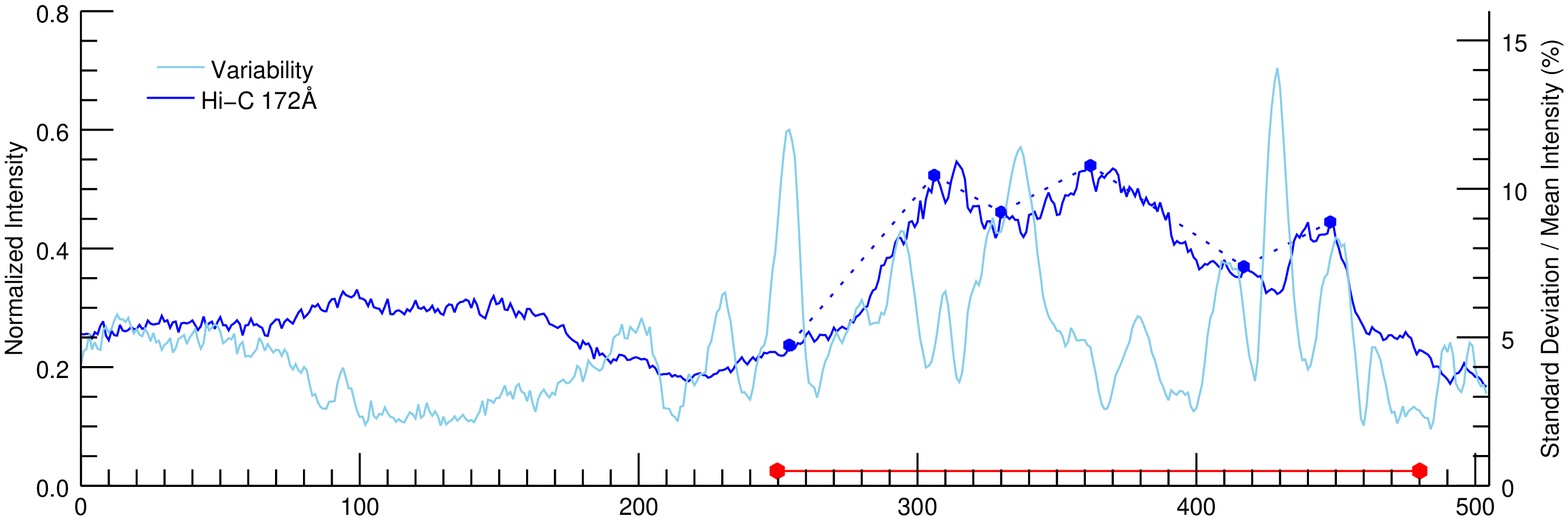}} %
  \caption{Time variability of features observed in the plage by Hi-C.
Normalized Hi-C intensity (dark blue) for the same slit position as in Fig. \ref{fig:fig9}. We highlight the bright structures in the plage (discussed in the text) with the blue dots. We measured the widths, lifetimes, and intensity variations of these features and they are similar to those of Fig. \ref{fig:fig8}. The sky blue line shows the ratio of the standard deviation of the intensities (in time) to the mean intensity (in time) expressed as a percentage. This measure shows a lower level of variability comparable to  that seen at lower spatial resolution in EIS observations, though over much shorter time-scales. The red dots show the locations of the boundaries of the plage for reference. These locations were chosen for the background/foreground subtraction. }
  \label{fig:fig10}
\end{figure*}

There is dynamic activity in the plage with similar intensity variations and lifetimes to that seen in Fig. \ref{fig:fig5} and Fig. \ref{fig:fig8}. This is shown in Fig. \ref{fig:fig10}. The widths of the brightening were more difficult to examine due to the smoothness of the emission in this slit position, but are similar to those measured above. The third slit position (S3) proved even more problematic. Most of the emission comes from the fan loops to the South, and only the edge of the plage region is actually in the outflow.

Our Hi-C analysis reveals that a similar systematic study of many outflow regions may be possible with the lower spatial resolution AIA data. Analysis of data from the first Hi-C flight found evidence that some loops observed by Hi-C showed evidence of substructure while others did not \citep{brooks_etal2013, peter_etal2013, delzanna_2013b}.  With hindsight it appears that the AIA spatial resolution may be good enough to separate the EIS outflow components; though the properties of the dynamic features we discuss would be less well constrained.

\subsection{Effect of the EIS Point-Spread-Function (PSF)}
The amorphous structure and relatively low EUV emission in the outflow regions make it difficult to identify any recognizable feature that contributes to the emission away from the plage area and bright fan loops. This raises an issue as to whether some part of the outflow might simply result from an instrumental effect, such as the spill-over of photons to adjacent pixels on the detector due to the EIS PSF.

The EIS optical performance was investigated on the ground prior to launch and the spatial resolution was measured to be close to the 2 arcsecond Nyquist limit expected for the detector pixel scale \citep{korendyke_etal2006}. On-orbit observations, however, suggest a lower spatial resolution. The smallest transition region brightenings detected by EIS have been used as point-like sources to estimate a value of 3--4 arcseconds by independent members of the EIS team (see the discussion on the EIS wiki at http://solarb.mssl.ucl.ac.uk/eiswiki/Wiki.jsp?page=TRbrighte\\
nings). Higher spatial resolution AIA images were also convolved with a Gaussian PSF to see what values would best reproduce the EIS raster images (see EIS Software note number 8 http://solarb.mssl.ucl.ac.uk/SolarB/eis\_docs/eis\_notes/08\_CO\\
MA/eis\_swnote\_08.pdf). That investigation found that a Gaussian PSF with a FWHM of 3--3.6 arcseconds produced the best match.

Some studies have also found evidence of an asymmetry and inclination to the PSF. It appears that in regions of strong intensity gradients the effect is to introduce a systematic shift in Doppler velocity (on the order of a few km s$^{-1}$) across the feature \citep{young_etal2012}. This was most dramatically demonstrated for EIS observations of a thin off-limb current sheet that developed following an X8.3 flare in September, 2017 \citep{warren_etal2018}. An asymmetric Gaussian PSF with a FWHM of 3 arcseconds in the X-direction and 4 arcseconds in the Y-direction was able to reproduce the reversed Doppler pattern observed around the current sheet. These values are consistent with measurements made during the 2012 Venus transit (Ugarte-Urra - private communication).

\begin{figure*}[t!]
  \centerline{%
    \includegraphics[width=1.0\textwidth]{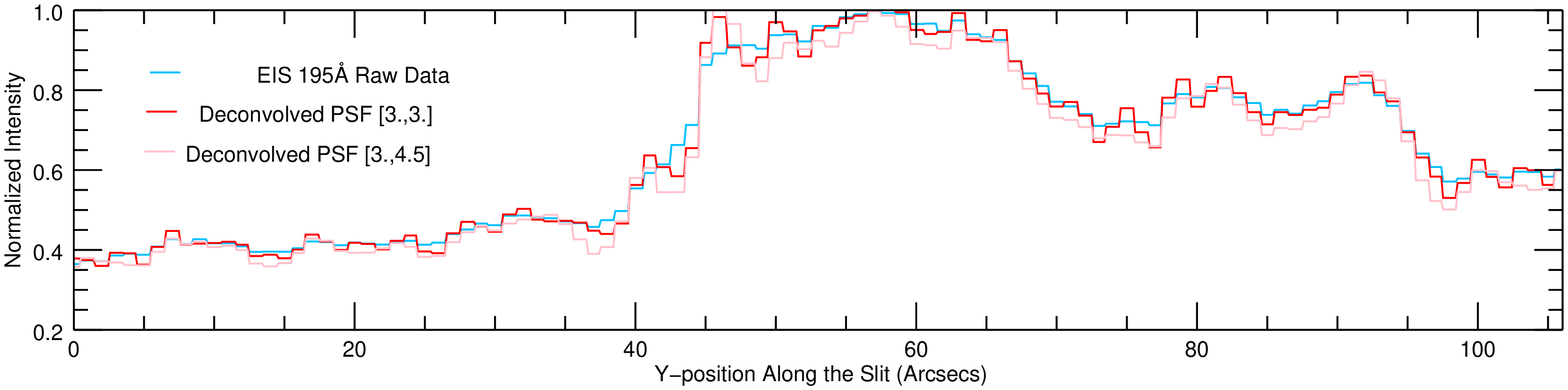}} %
\caption{EIS raw data and two PSF deconvolutions. The intensity distribution for the same slit position as shown in Fig. \ref{fig:fig5} compared to two simulations of the data that attempt to deconvolve the EIS PSF. We show the Fe XII 195.119\,\AA\, raw intensities in blue, and the deconvolution using a symmetric Gaussian PSF with a full-width-half-maximum (FWHM) of 3 arcseconds. We also show a deconvolution using an asymmetric Gaussian PSF with a FWHM of 3 arcseconds in the X-direction and 4.5 arcseconds in the Y-direction.}
  \label{fig:fig11}
\end{figure*}

We do not expect spurious velocity artifacts to affect observations in large homogenous areas of outflow such as we have analyzed here. We have, however, investigated the effect of a Gaussian and asymmetric Gaussian PSF on the distribution of intensities through the outflow region along the slit position analyzed in Fig. \ref{fig:fig5}. We show the results of this experiment in Fig. \ref{fig:fig11}. The figure contrasts the raw intensity distribution (blue line) with the intensity distribution produced after deconvolution using a Gaussian PSF with a FWHM of [3,3] arcseconds (red line) and an asymmetric Gaussian PSF with a FWHM of [3,4.5] arcseconds (pink line). The raw and deconvolved data do not show any significant differences. Although small deviations are introduced by the PSF, the intensities remain highly correlated, and the general features we are interested in (the plage and background) are largely unaffected. The modeled PSF implies that the peak intensity in a feature should drop to its FWHM within 3--4.5 arcseconds, whereas the background component is slowly changing and the standard deviation from the mean is less than 10\% over the lower 0--40 arcseconds of the slit.

\subsection{Magnetic topology model}
For magnetic topology modeling (see Section \ref{discussion}) we use a potential field source surface (PFSS) extrapolation made with the package distributed in SolarSoftware \citep{schrijver&derosa_2003}. The package traces field lines from potential field models archived at 6hr cadence for specified Carrington longitudes and heliographic latitudes. AR 12712 was on disk during Carrington rotation 2204 and appropriate ranges to cover the region were Carrington longitudes [108.6,258.6] and heliograhic latitudes [-61.0,59.0]. The PFSS model extrapolates up to 2.5\,R$_{\sun}$. To illustrate the global structure and magnetic configuration of the active region we drew a selection of 275 field lines on top of a magnetogram obtained by the Helioseismic and Magnetic Imager (HMI) on SDO. The selected field lines start from radial distances within a range of 1.02--1.5\,R$_{\sun}$. The HMI magnetogram was downloaded from the Joint Science Operations Center (JSOC) at Stanford University. It is a level 1 full disk magnetogram obtained during the Hi-C flight at 18:56UT.
\section{discussion}
\label{discussion}

\begin{figure*}[t!]
  \centerline{%
    \includegraphics[width=0.32\textwidth]{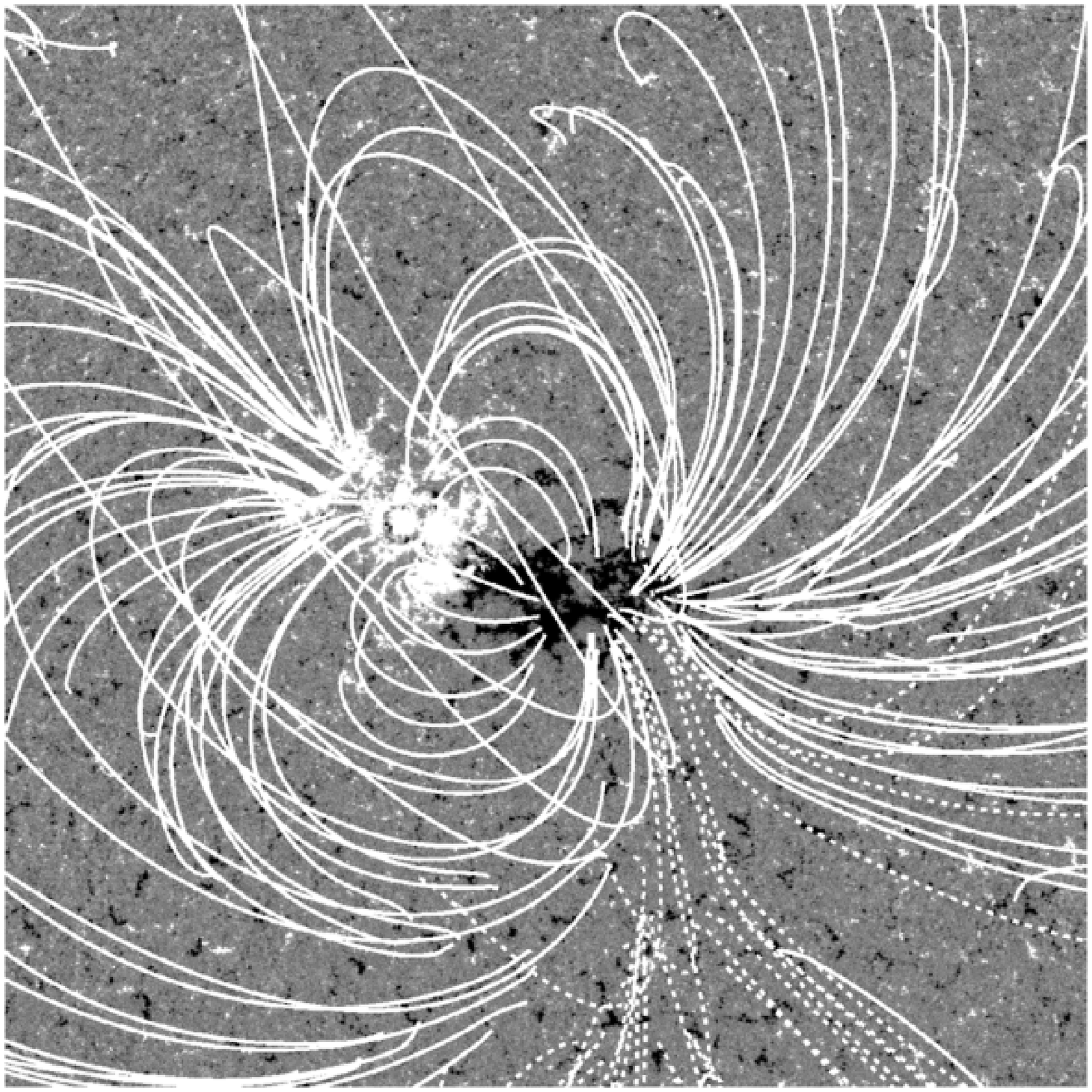} %
    \includegraphics[width=0.32\textwidth]{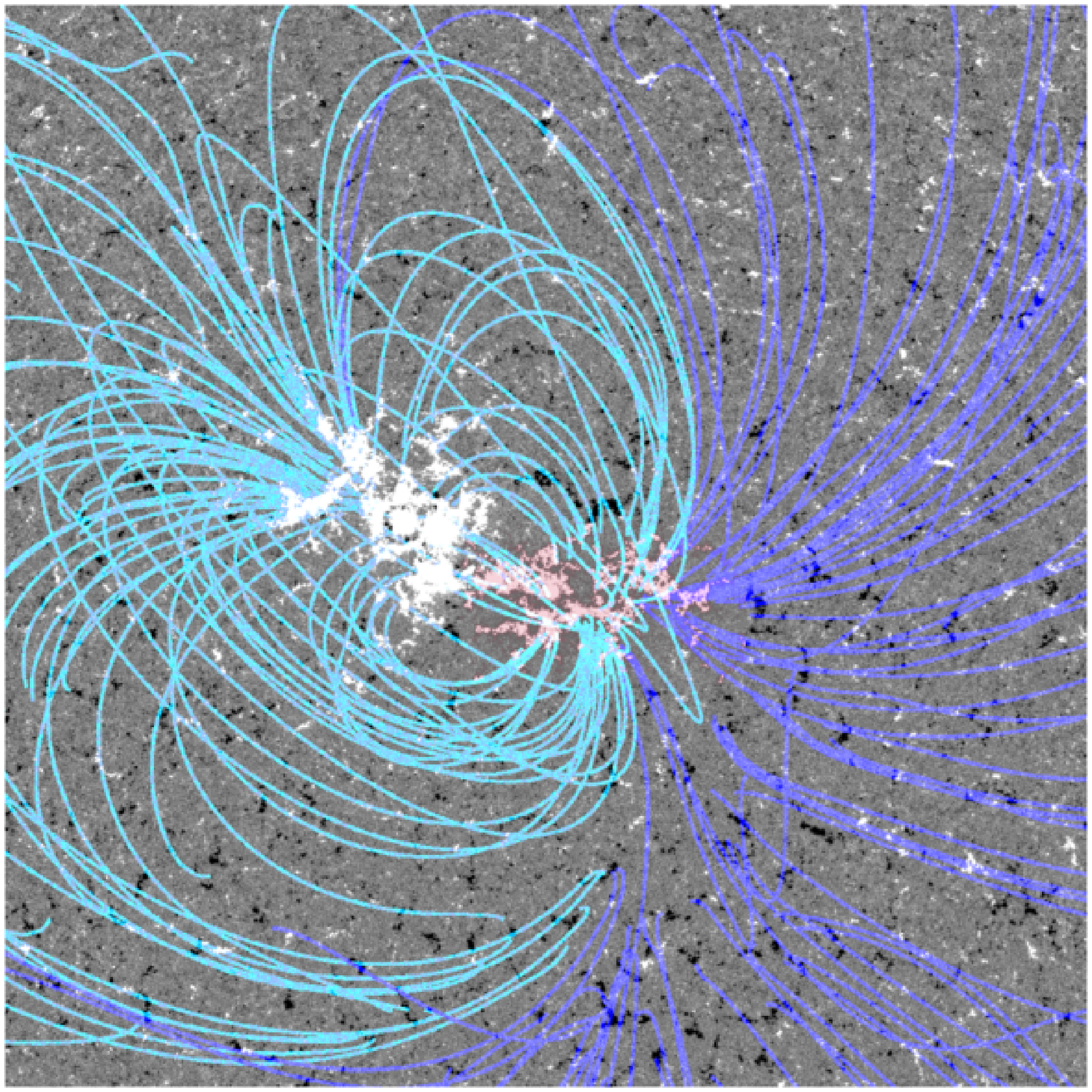} %
    \includegraphics[width=0.15\textwidth]{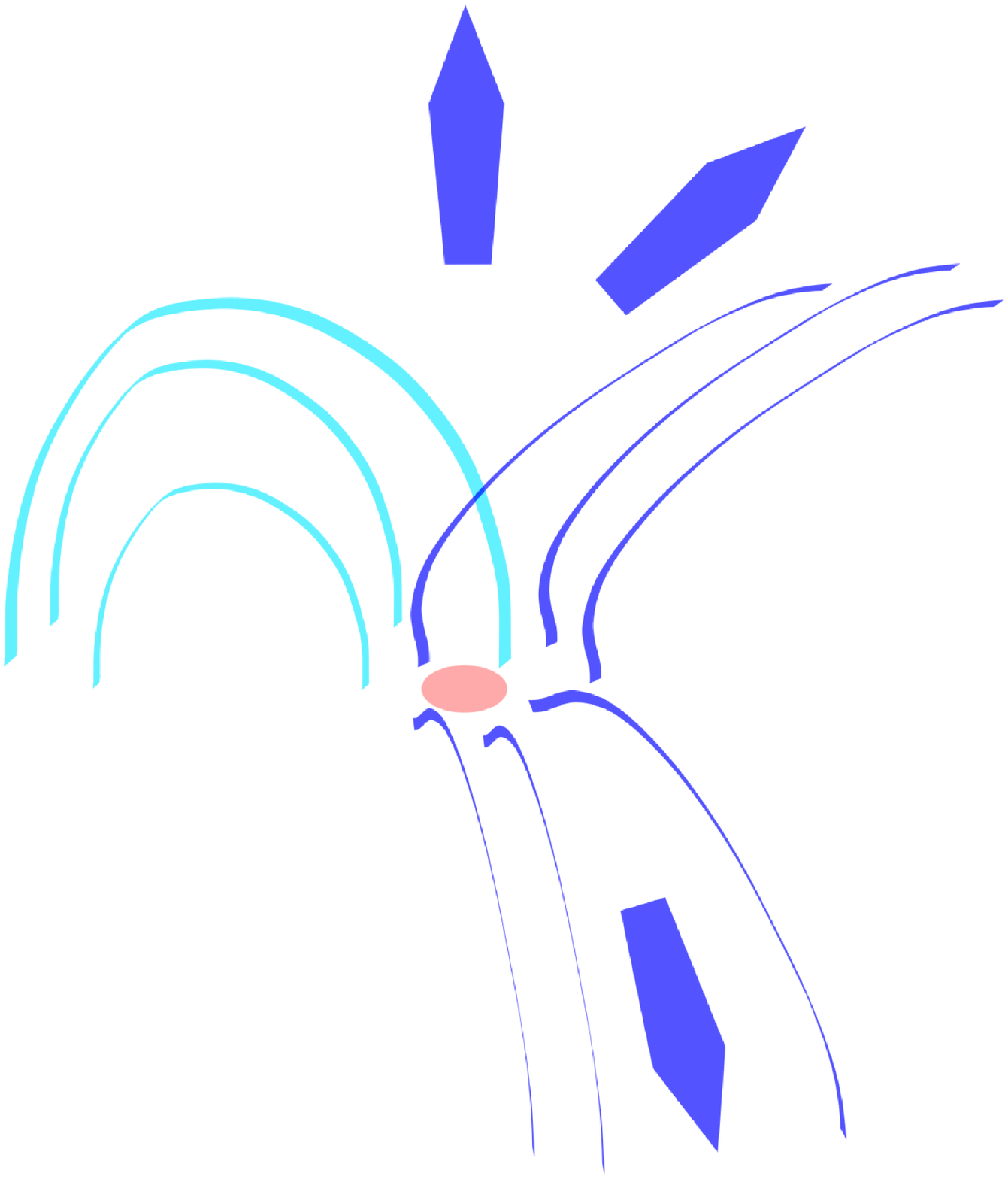} %
    \includegraphics[width=0.15\textwidth]{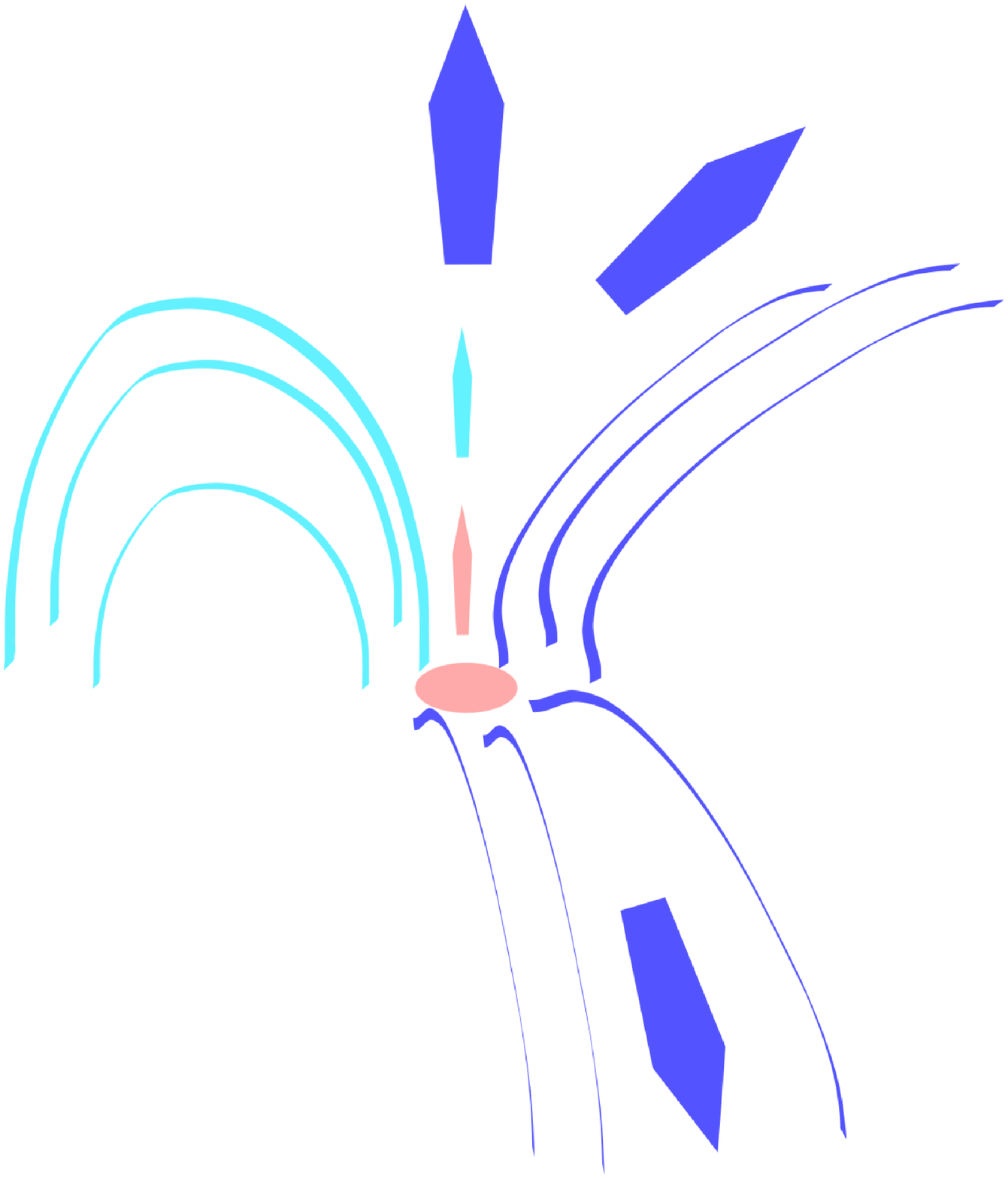}} %
  \caption{Description of the magnetic field and activity in the EIS/Hi-C observations
Left panel: PFSS extrapolation of an HMI magnetogram taken at 18:56UT on 2018, May 29. We show a selection of extrapolated magnetic field lines to illustrate the magnetic configuration of the region. The white/black patches represent positive/negative polarities. Closed/open field lines are represented by solid/dashed white lines. Middle panel: a colored version of the left panel representing the general structure of the active region and illustrating the features shown in Fig. \ref{fig:fig5}. We show closed loops with a coronal composition in sky blue, and open and closed long field lines where upflows are measured by EIS in blue. Pink represents the plage areas with a photospheric composition. The field lines are randomly selected when generating the figure so are slightly different from those in the left panel. Right panel: a schematic cartoon drawn to qualitatively explain the features shown in this Figure, Fig. \ref{fig:fig5}, and the EIS measurements. All field lines extend from positive polarities and converge in the plage area at the base of the outflows observed by EIS (left cartoon). As an example, component reconnection can take place between the closed sky blue loops and the blue open and closed long field lines. This process can release material into the outflow from both the pink plage region and the sky blue closed loops (right cartoon). The colors of the arrows represent the sources of the flows. }
  \label{fig:fig12}
\end{figure*}

Our outflow composition measurements, aided by high spatial resolution Hi-C images, identify two drivers of the outflows that went undetected at lower spatial resolution. Previous Hi-C observations show the presence of wave motions in both active region loops and moss \citep{morton_Etal2013,morton_etal2014}, and theoretical knowledge of the FIP effect based on magnetohydrodynamic (MHD) wave models \citep{laming_2004,laming_2015} provide a framework to understand our results. First, theory and observations of evolving active regions suggest that plasma needs to be confined for some time (at least several hours) before the plasma composition becomes enhanced \citep{widing&feldman_2001}. So the coronal component of the outflow is likely to be a signature of plasma that has been confined in closed magnetic loops after emergence, perhaps in the active region core, and then released into the outflow when these loops open. Second, dynamic activity in the upper transition region in the plage studied here occurs on time-scales (a few minutes) that modeling suggests are too short for the FIP effect to take place. Plasma is ejected rapidly through the region where the FIP effect is assumed to operate (the upper chromosphere), and injected directly into the outflow from the plage region with the observed unenhanced (photospheric) composition. As suggested by the coronal-contraflow model \citep{mcintosh_etal2012}, there may be a draining phase detectable as down flowing plasma at lower temperatures. We have deliberately attempted to spatially isolate the upflow components here, since they are the only ones that produce plasma that can contribute to the solar wind.

The magnetic configuration of AR 12712 suggests how this picture could apply here (Fig. \ref{fig:fig12}). The closed AR core loops (sky blue in the Figure) extend from positive polarities and converge in the negative polarity area where plage (pink region) is located at the base of the outflows observed by EIS (blue). Open field lines and long loops that close distant to the AR (blue in the Figure) also converge in this negative polarity region. As an example, the schematic cartoon in Fig. \ref{fig:fig12} illustrates the process. Component reconnection can take place between the core loops and the open or closed long field lines. This can release material into the outflow from both the plage region (pink arrow) and the closed loops (sky blue arrow). The photospheric and coronal composition plasma then escapes and expands into the outflow (blue arrows). 

This scenario can potentially explain the variability of the slow wind composition observed in-situ. In this active region, the outflows show a coronal composition at the spatial scales measured by EIS, but a critical point is that plasma with a photospheric composition is also being injected even though it went undetected until now. The coronal component dominates in this case (only about 1/3 of the emission between the red dots in Fig. \ref{fig:fig5} is being contributed by the plage at the formation temperature of 195.119\,\AA; 1.6\,MK), but this likely depends on the time, location, and active region. It is not hard to imagine a scenario where the photospheric component is the dominant contributor and the total outflow signature is photospheric.

\appendix

As discussed in Section \ref{velocity}, we applied a non-standard additional correction to remove a residual orbital drift that
remained in the data after following the standard correction procedures. 
The standard correction is based on a neural network calculation performed early in the mission. Updates were made following significant instrument configuration changes (e.g. slit and grating focus) but have not been made since 2008. So we often find that a residual orbital drift remains that needs further correction. After assessing the accuracy of the orbital drift correction for the 
observations used in this paper, we found that in fact this was true of the datasets presented in this article. The residual drift was removed by averaging the velocities in the Y direction in the upper 100 pixels of the CCD, smoothing the resultant curve over 5 pixels in the X direction, and subtracting this smoothed velocity function from the data.

\begin{figure}[t!]
  \centerline{%
    \includegraphics[width=0.5\textwidth]{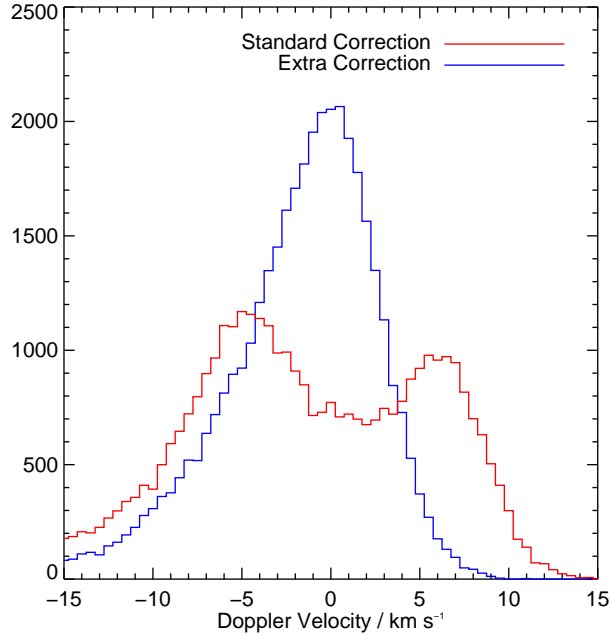}} %
  \caption{ Distribution of Doppler velocities derived from the Fe XIII 202.044\,\AA\, spectra taken at 14:54:11UT on 2018, May 29.
The velocity map for these data is shown in the 4th panel of the lower row of Fig. \ref{fig:fig1}. The red histogram shows the 
measurements made using the standard orbital drift correction available in Solar Software \citep{kamio_etal2010}. The blue histogram
shows the measurements after out additional correction.
}
  \label{fig:figA1}
\end{figure}

We illustrate the improvement in measurements achieved by this method in Fig. \ref{fig:figA1}. The red histogram shows the results after
following the standard procedure. The Doppler velocity distribution is approximately bimodal, indicating a shift
across the FOV with an amplitude of 10.5\,km s$^{-1}$ (the drift is red to blue in this case). The blue histogram shows the results
after correcting for the residual drift. The apparent bimodality is removed, and we are left with a single peaked distribution with a
standard deviation of $\sim$3.5\,km s$^{-1}$. This is comparable to the rms error of 4.4\,km s$^{-1}$ achived by the standard method
when applied to data earlier in the mission.

\acknowledgments 
D.H.B. thanks Ignacio Ugarte-Urra for discussions on the EIS PSF. The work of D.H.B. was performed under contract with the Naval Research Laboratory and was funded by the NASA \textit{Hinode} program. S.K.T. gratefully acknowledges support by NASA contracts NNG09FA40C (\textit{IRIS}), and NNM07AA01C (\textit{Hinode}). We acknowledge the High-resolution Coronal Imager (Hi-C 2.1) instrument team for making the second re-flight data available under NASA Heliophysics Technology and Instrument Development for Science (HTIDS) Low Cost Access to Space (LCAS) program (proposal HTIDS17\_2-0033). MSFC/NASA led the mission with partners including the Smithsonian Astrophysical Observatory, the University of Central Lancashire, and Lockheed Martin Solar and Astrophysics Laboratory.  Hi-C 2.1 was launched out of the White Sands Missile Range on 2018 May 29. \textit{Hinode} is a Japanese mission developed and launched by ISAS/JAXA, with NAOJ as domestic partner and NASA and STFC (UK) as international partners. It is operated by these agencies in co-operation with ESA and NSC (Norway). We thank the ACE SWICS and SWEPAM instrument teams and the ACE Science Center for providing the ACE data. The AIA data used are courtesy of NASA/SDO and the AIA, EVE, and HMI science teams. CHIANTI is a collaborative project involving George Mason University, the University of Michigan (USA) and the University of Cambridge (UK).


\end{document}